\documentclass[12pt]{article}

\usepackage{graphicx}
\usepackage{cite}
\usepackage{amsmath,amssymb,amsthm,bm}
\usepackage{physics}
\usepackage{mathtools}
\usepackage{geometry}
\usepackage{hyperref}

\makeatletter
\@addtoreset{equation}{section}
\makeatother

\newcommand{\ii}{\mathrm{i}}
\newcommand{\Geff}{W^{(1)}}

\begin{document}


\begin{titlepage}
 

\begin{Large}
\vspace{10mm}
\begin{center}
{\bf The Fuzzy-Sphere as a Black Hole in \\ the IKKT Matrix Model: \\ An Assessment}
\end{center}
\end{Large}
\vspace{10mm}


\begin{center}
 Koichiro M{\sc atsumoto}$^{1), 2)}$
 \footnote{E-mail address: matsumoto.koichiro@musashi.ed.jp}\\
 \vspace{5mm}
 $^{1)}$
 {\it Musashi High School and Junior High School}\\
 {\it Nerima, Tokyo 176-8535, Japan}\\
 $^{2)}$
  {\it Waseda Junior and Senior High School}\\
 {\it Shinjuku, Tokyo 162-8654, Japan}\\
\end{center}
\vspace{5mm}

 
\begin{abstract}
Black-hole thermodynamics has been reproduced with remarkable success
from the BFSS matrix model, but the corresponding test in the closely
related IKKT matrix model has awaited its recently proposed
finite-temperature formulation, within which we investigate this
correspondence.
Introducing a cubic Myers term, we compute the complete one-loop
thermodynamics -- free energy, internal energy, entropy, and heat
capacity -- of a fuzzy $S^2$ background, and compare against the D0
and D2 black holes as the closest available reference points, since
no gravitational dual is known for this specific background.
At fixed Myers coupling, the leading entropy exhibits an $O(N^3)$
scaling rather than the $O(N^2)$ behavior characteristic of conventional
black holes. Holding the physical radius fixed instead yields a genuine
$O(N^2)$ scaling, showing that the power of $N$ depends on the large-$N$
prescription and is therefore not by itself an unambiguous test of a
black-hole interpretation. We further find no intrinsic Hawking
temperature -- only a characteristic scale -- and a heat capacity that
is negative at low temperature but positive at high temperature.
The absence of an intrinsic Hawking temperature, tied to the missing
gravitational dual rather than to the choice of large-$N$ prescription,
persists in either scaling limit. Rather than establishing a
black-hole interpretation of the finite-temperature IKKT matrix model,
these results provide a quantitative benchmark against which future
analytical, numerical, and holographic investigations can be tested.
\end{abstract}

\end{titlepage}

\section{Introduction}

The IKKT matrix model was proposed as a non-perturbative
formulation of type IIB superstring theory in the light-cone gauge,
arising as a matrix regularization of the Green-Schwarz superstring
\cite{ishibashi_kawai_kitazawa_tsuchiya-1997}. The basic structure and
symmetries of the model were further elucidated in
\cite{aoki_iso_kawai_kitazawa_tsuchiya_tada-1999,aoki_iso_kawai_kitazawa_tada-1998}.
In this framework spacetime itself is not put in by hand but emerges
dynamically from the eigenvalue distribution of the bosonic matrices.
This emergent-spacetime scenario motivated an extensive early research
program: in the Lorentzian version of the model, Monte Carlo studies
established the mechanism of a spontaneous breaking of rotational
symmetry \cite{nishimura_vernizzi-2000,anagnostopoulos_nishimura-2002},
and subsequent simulations demonstrated that this mechanism dynamically
generates an expanding (3+1)-dimensional spacetime out of the
ten-dimensional model \cite{kim_nishimura_tsuchiya-2012}, while in
the Euclidean version, the Gaussian expansion method provided analytic
evidence for a spontaneous breaking of the ten-dimensional rotational
symmetry down to four dimensions
\cite{nishimura_sugino-2002,kawai_kawamoto_kuroki_matsuo_shinohara-2002,kawai_kawamoto_kuroki_shinohara-2003}.
This body of work makes the model a natural setting for probing
non-perturbative aspects of quantum gravity, in particular the origin
of spacetime. A genuine formulation of quantum gravity should also
account for black holes, among the most important non-perturbative
gravitational objects, and in particular for their thermodynamic
properties.

Black hole thermodynamics has long occupied a central place in the
study of quantum gravity. Bekenstein proposed that a black hole
carries an entropy proportional to its horizon area
\cite{bekenstein-1973}. This was placed on a firmer footing by the
formulation of four laws of black-hole mechanics directly paralleling
the laws of thermodynamics \cite{bardeen_carter_hawking-1973}, by
Hawking's derivation of the associated thermal radiation
\cite{hawking-1975}, and by the Euclidean path-integral formulation of
the partition function from which such thermodynamic quantities can be
computed~\cite{gibbons_hawking-1977}. Together these established
that black holes behave as genuine thermodynamic systems, raising the
question of what microscopic degrees of freedom account for this
entropy. String theory provided the first controlled answer to this
question: Strominger and Vafa showed that the Bekenstein--Hawking
entropy of a class of extremal black holes can be reproduced by
counting BPS bound states of D-branes \cite{strominger_vafa-1996},
establishing a direct link between black-hole thermodynamics and the
microscopic, non-perturbative degrees of freedom of string theory. A
complementary, non-supersymmetric approach emerged shortly afterward
within Matrix theory itself, where the mass-entropy relation and
Hawking temperature of Schwarzschild black holes were reproduced,
up to numerical coefficients, directly from the thermodynamics of the
BFSS matrix model \cite{banks_fischler_klebanov_susskind-1998,ohta_zhou-1998}. 
This success motivates the broader program of testing whether other
non-perturbative formulations of string/M-theory reproduce black hole
thermodynamics from first principles.
It is therefore natural to ask whether the same program can be carried out 
within the IKKT matrix model, which is expected to provide 
a non-perturbative definition of type IIB superstring theory.

This expectation has already been tested with remarkable success in the
closely related BFSS matrix model
\cite{banks_fischler_shenker_susskind-1997}, which conjectures a
duality between D0-brane quantum mechanics and M-theory (equivalently,
type IIA supergravity in the appropriate limit). This duality is a
non-conformal instance of the more general correspondence between the
large-$N$ limit of maximally supersymmetric gauge theories and
supergravity on the near-horizon geometry of the corresponding branes
\cite{maldacena-1997,itzhaki_maldacena_sonnenschein_yankielowicz-1998},
with the free energy and entropy of the dual near-extremal black
D0-brane background computed on the gravity side in
\cite{klebanov_tseytlin-1996}.

Ahead of the lattice studies, an
analytic, non-perturbative approach based on a Gaussian (gap-equation)
approximation to the strongly coupled quantum mechanics was developed
and used to compute black-hole thermodynamic
quantities~\cite{lowe-1998,kabat_lifschytz-2000,
kabat_lifschytz_lowe-2001a,kabat_lifschytz_lowe-2001b}. A series of
lattice Monte Carlo studies established this correspondence step by
step: early tests of the
low-temperature black-hole regime
\cite{catterall_wiseman-2008,anagnostopoulos_hanada_nishimura_takeuchi-2008}
were followed by an extraction of the Schwarzschild radius from the
Wilson loop \cite{hanada_miwa_nishimura_takeuchi-2009} and a
confirmation of the leading higher-derivative correction to the
black-hole free energy \cite{hanada_hyakutake_nishimura_takeuchi-2009},
culminating in a high-precision confirmation of the predicted internal
energy \cite{hanada-2014}, extended to the continuum limit and to
finite $N$~\cite{hanada_hyakutake_ishiki_nishimura-2016}, and most
recently extended to unprecedented
accuracy via the BMN deformation \cite{pateloudis_et_al-2023}.

Together
these results have confirmed the gauge/gravity correspondence between
the thermal matrix model and the dual black-hole geometry to good
precision. In the IKKT matrix model, by contrast, the corresponding
relation between the matrix model and black-hole thermodynamics
remains far less understood, largely because a finite-temperature
formulation of the model --- required to even pose the question --- has
only recently become available.

This obstacle was recently addressed by
Ref.~\cite{laliberte_brahma-2023}, which introduced a finite-temperature
deformation of the Euclidean IKKT matrix model by compactifying the
Euclidean time matrix on a circle of circumference $\beta=T^{-1}$, with
anti-periodic boundary conditions on the fermionic matrix, in close
analogy with the standard thermal compactification of the target-space
time direction in superstring theory. This construction provides, for
the first time, a concrete framework in which thermodynamic
quantities --- free energy, internal energy, entropy --- can in
principle be computed directly from the IKKT matrix model at finite temperature. 
Although this construction makes thermodynamic observables well defined,
the correspondence between such thermodynamic quantities and the expected
properties of black holes in the dual gravitational description has not yet
been established: no explicit
one-loop computation of the finite-temperature effective action around
a nontrivial classical background, from which these thermodynamic
quantities could be extracted and compared against gravitational
expectations, has so far been carried out. A related but distinct
approach, extending the same mass-generation mechanism to the BFSS
model and constructing a black-hole solution whose entropy arises from
fermionic excitations on a fuzzy-sphere background, has recently been
proposed in Ref.~\cite{laurenzano_wheater-2025}.

In this work we compute this one-loop effective action explicitly. We
stabilize the classical background by deforming the Euclidean IKKT
action with the standard cubic Myers dielectric interaction, which
polarizes the $N$ matrix degrees
of freedom into a bound state supported on a fuzzy two-sphere $S^2$
\cite{madore-1992}, fixing the on-shell value of the non-commutativity
scale $\alpha$ independently of $N$ (the physical radius itself grows
with $N$ at fixed Myers coupling, Sec.~\ref{sec:fuzzyS2}). A
deformation of the IKKT matrix model along these lines was first
discussed in Ref.~\cite{bonelli-2002}.
This background provides the simplest nontrivial classical solution 
of the deformed model and serves as a natural testing ground for finite-temperature dynamics.
Expanding the action to quadratic order in
fluctuations about this background and integrating out the bosonic and
fermionic fluctuation modes, we obtain the one-loop effective action at
finite temperature, from which the free energy, internal energy, and
entropy of the system follow directly.

Comparing the resulting thermodynamic quantities with those expected
from a dual black-hole geometry, we identify several nontrivial discrepancies.
Most notably, the entropy scales as $O(N^3)$, 
in contrast to the conventional $O(N^2)$ behavior of black-hole entropy.
We also find that the heat capacity changes sign between low and high
temperature (negative, then positive), so that no Hawking-temperature
coefficient can be fixed from the matrix-model calculation alone. These issues are discussed in
detail in Sec.~\ref{sec:tests}.
Rather than viewing these discrepancies as evidence against the
gauge/gravity correspondence, we regard them as important clues
toward identifying the appropriate gravitational dual and the role
of corrections beyond the present one-loop approximation. These
findings clarify the nature of the mismatch between black-hole
thermodynamics and the finite-temperature IKKT matrix model; we discuss
possible sources of the mismatch and directions for resolving it in
Sec.~\ref{sec:discussion}.

The remainder of this paper is organized as follows. In
Sec.~\ref{sec:thermal} we review the finite-temperature compactification
of the Euclidean IKKT matrix model. In Sec.~\ref{sec:oneloop} we
construct the fuzzy $S^2$ background stabilized by the Myers term and
compute the corresponding one-loop effective action. In
Sec.~\ref{sec:thermo} we derive the free energy, entropy, internal
energy, and heat capacity from this effective action. In
Sec.~\ref{sec:tests} we confront these thermodynamic quantities with
the properties expected of a dual black hole. We conclude in
Sec.~\ref{sec:discussion} with a discussion of the implications of our
results and directions for future work. 
Technical details of the background-field expansion and 
the Matsubara sums are collected in the Appendices.

\section{Finite-temperature compactification}
\label{sec:thermal}

In this section we review the construction of
Ref.~\cite{laliberte_brahma-2023}, who introduced a finite-temperature
deformation of the Euclidean IKKT matrix model by compactifying the
Euclidean time matrix on a circle, in close analogy with the standard
thermal compactification of the target-space time direction in
superstring theory.\footnote{Our normalization of the circumference of
the thermal circle, $\beta=T^{-1}$, differs from that of
Ref.~\cite{laliberte_brahma-2023}, where the circumference is taken to
be $2\pi\beta_{\rm LB}$; the two are related by $\beta=2\pi\beta_{\rm LB}$.}

The Euclidean IKKT action is given by
\begin{align}
  I_E = -\frac{1}{4g^2}\,\mathrm{Tr}[A_\mu, A_\nu]^2
        -\frac{1}{2g^2}\,\mathrm{Tr}\!\left(\bar\psi\,\Gamma^\mu[A_\mu,\psi]\right),
  \label{eq:ikkt_euclidean}
\end{align}
where $\mu,\nu=0,\dots,9$, $A_\mu$ and $\psi$ are $N\times N$ Hermitian matrices,
with $A_\mu$ being bosonic and $\psi$ a ten-dimensional Majorana--Weyl spinor,
$\bar\psi \equiv \psi^\dagger\Gamma^0$ is the Dirac conjugate,
$g$ is the coupling constant,
and indices are contracted with the Euclidean metric $\delta_{\mu\nu}$.
Unlike Ref.~\cite{laliberte_brahma-2023}, who obtain \eqref{eq:ikkt_euclidean}
by Wick-rotating a Lorentzian action, we take it directly as our starting
point in Euclidean signature, working throughout with the Euclidean
Majorana--Weyl fermion $\psi$ and its Dirac conjugate $\bar\psi$.

We introduce finite temperature
by compactifying the Euclidean time direction on a circle of circumference
$\beta$, where $\beta = T^{-1}$ is the inverse temperature, and imposing
anti-periodic boundary conditions on the fermionic matrix --- the matrix
analogue of the standard prescription for obtaining a thermal state of the
superstring by compactifying its Euclidean target-space time direction with
anti-periodic fermions (a Scherk--Schwarz-type mechanism~\cite{scherk_schwarz-1979}).
The compactification is implemented through the unitary translation operator
\begin{align}
  U = \mathbf{1}_N \otimes e^{-\ii2\pi q}\,e^{-\ii p},
  \label{eq:unitary_U}
\end{align}
where $p$ and $q$ satisfy the Heisenberg algebra $[q,p]=\ii$.
The thermal boundary conditions are imposed algebraically as
\begin{align}
  U^{-1}A_0\,U &= A_0 + \beta, \label{eq:bc_A0}\\
  U^{-1}A_i\,U &= A_i,         \label{eq:bc_Ai}\\
  U^{-1}\psi\, U &= -\psi,     \label{eq:bc_psi}
\end{align}
so that $A_i$ ($i=1,\dots,9$) are periodic (bosonic) and $\psi$ is
anti-periodic (fermionic).

The Heisenberg algebra $[q,p]=\ii$ fixes the mode numbers entering the
general solution of \eqref{eq:bc_A0}--\eqref{eq:bc_psi} (a point only
briefly indicated in the original construction): conjugation by $U$
acts as $e^{\ii np}\mapsto e^{\ii np}$ for integer $n$ and
$e^{\ii rp}\mapsto-e^{\ii rp}$ for half-integer $r$ (the phase
$e^{\ii2\pi q}$ in \eqref{eq:unitary_U} realizes the fermion-parity
operator $(-1)^F$), together with $\beta q\mapsto\beta q+\beta$,
reproducing \eqref{eq:bc_A0}--\eqref{eq:bc_psi} with the mode expansions
\begin{align}
  A_0 &= \beta q\otimes\mathbf{1}_N + \sum_{n\in\mathbb{Z}} A_0^n\otimes e^{\ii np},
  \label{eq:mode_A0}\\
  A_i &= \sum_{n\in\mathbb{Z}} A_i^n\otimes e^{\ii np},
  \label{eq:mode_Ai}\\
  \psi &= \sum_{r\in\mathbb{Z}+\frac{1}{2}} \psi^r\otimes e^{\ii rp},
  \label{eq:mode_psi}
\end{align}
where $A_\mu^n,\,\psi^r\in\mathrm{Mat}_N$ are $N\times N$ matrices
and $e^{\ii np},\,e^{\ii rp}$ act on the Hilbert space $\mathcal{H}$.
The full trace decomposes as $\mathrm{Tr}=\mathrm{Tr}_N\otimes\mathrm{Tr}_{\mathcal{H}}$.

This construction admits a geometric interpretation as a periodic array
of mirror D-instanton configurations along the $A_0$ direction, though
it is not needed for the computation that follows.

Two identities play a central role in the computation.
First, from the Heisenberg algebra one derives
\begin{align}
  [q,\,e^{\ii np}] = -n\,e^{\ii np},
  \label{eq:q_einp}
\end{align}
which shows that the adjoint action of the zero-mode part of $A_0$ on the
$n$-th bosonic mode generates the Matsubara frequency $\omega_n = \beta n$,
and similarly $\omega_r = \beta r$ for the fermionic modes.
Second, the Hilbert-space trace satisfies the orthogonality relation
\begin{align}
  \mathrm{Tr}_{\mathcal{H}}(e^{\ii(n+m)p}) = V\,\delta_{n+m,\,0},
  \label{eq:orthogonality}
\end{align}
where $V=\int_{-\infty}^{\infty}dq$ is the (infinite) volume factor of $\mathcal{H}$,
independent of $N$.

Using \eqref{eq:q_einp} and \eqref{eq:orthogonality} together with the
standard tensor-product commutator expansion, one evaluates each term
in $I_E$ and performs the trace.
After combining the cubic interaction terms via the cyclicity of $\mathrm{Tr}_N$,
the mode-expanded action reads
\begin{align}
  I_E = \frac{V}{2g^2}\,\mathrm{Tr}_N\Biggl[
    &\sum_n(\beta n)^2 \delta^{ij}A_i^{-n}A_j^n
    +2\sum_{n,m}\beta(n+m)\delta^{ij}[A_0^n,A_i^m]A_j^{-n-m}
    \notag\\
    &-\frac{1}{2}\sum_{n,m,n'}\delta^{ij}[A_0^n,A_i^m][A_0^{n'},A_j^{-n-m-n'}]
    \notag\\
    &-\frac{1}{2}\sum_{n,m,n'}\delta^{ik}\delta^{jl}[A_i^n,A_j^m][A_k^{n'},A_l^{-n-m-n'}]
    \notag\\
    &+\sum_r\beta r\,\bar\psi^{-r}\Gamma^0\psi^r
    -\sum_{n,r}\bar\psi^{-n-r}\Gamma^\mu[A_\mu^n,\psi^r]
  \Biggr].
  \label{eq:mode_expanded_action}
\end{align}
The mode-expanded action \eqref{eq:mode_expanded_action} serves as the
starting point for the background-field expansion around the Fuzzy $S^2$
solution carried out in Sec.~\ref{sec:oneloop}.

As a consistency check on \eqref{eq:mode_expanded_action}, consider the
decompactification limit $\beta\to\infty$. The mass-type terms
$(\beta n)^2$ and $\beta r$ send all non-zero Matsubara modes
$A_\mu^{n\neq0}$ to infinite mass, decoupling them; since the fermion sector
has no $r=0$ mode (the spectrum being strictly half-integer), the fermionic
contribution vanishes identically. Only the bosonic zero modes survive, and
\eqref{eq:mode_expanded_action} reduces to
\begin{align}
  I_E \ \xrightarrow{\ \beta\to\infty\ }\ 
  -\frac{V}{4g^2}\,\mathrm{Tr}_N[A_i^0,A_j^0]^2,
  \label{eq:decompactification}
\end{align}
the zero-temperature bosonic IKKT action for the zero-mode matrices, as
expected.

\section{One-loop effective action on the Fuzzy \texorpdfstring{$S^2$}{S\^{}2}}
\label{sec:oneloop}

\subsection{The fuzzy \texorpdfstring{$S^2$}{S\^{}2} background}
\label{sec:fuzzyS2}

The fuzzy two-sphere~\cite{madore-1992} is the standard non-commutative
regularization of the ordinary sphere $S^2$: its algebra of functions is
truncated to the finite-dimensional space of $N\times N$ Hermitian
matrices, realized concretely via the spin-$j$ ($j=(N-1)/2$) irreducible
representation matrices $L_i$ ($i=1,2,3$) of $\mathrm{SU}(2)$,
\begin{align}
  [L_i,L_j]=\ii\varepsilon_{ijk}L^k, \qquad L^iL_i\equiv L^2=C_2\mathbf{1}_N,
  \qquad C_2=j(j+1).
  \label{eq:su2gens}
\end{align}
Embedding coordinates $X_i=\alpha L_i$ for a real scale $\alpha$ satisfy
the defining relation of a sphere of radius $R=\alpha\sqrt{C_2}$,
\begin{align}
  X^iX_i = \alpha^2L^2 = \alpha^2C_2\,\mathbf{1}_N \equiv R^2\mathbf{1}_N,
\end{align}
but with non-vanishing, $O(N^{-1})$ commutators $[X_i,X_j]=\ii\alpha
\varepsilon_{ijk}X^k$ in place of the vanishing commutators of ordinary
$\mathbb{R}^3$ coordinates restricted to $S^2$: the fuzzy sphere reduces
to an honest commutative $S^2$ of fixed radius $R$ only in the combined
limit $N\to\infty$, $\alpha\to0$ with $\alpha\sqrt{C_2}=R$ held fixed, so
that $1/N$ plays the role of a non-commutativity parameter. This
construction is the standard building block for
non-commutative field theories and matrix models, and appears here as the
kind of curved, compact background we want to realize as a genuine
classical solution of the IKKT matrix model.

The Euclidean IKKT action $I_E=I_B+I_F$ of \eqref{eq:ikkt_euclidean},
split as in \eqref{eq:SBSFdef} into
\begin{align}
  I_B \equiv -\frac{1}{4g^2}\mathrm{Tr}[A_\mu,A_\nu]^2, \qquad
  I_F \equiv -\frac{1}{2g^2}\mathrm{Tr}\!\left(\bar\psi\,\Gamma^\mu[A_\mu,\psi]\right),
  \label{eq:SBSFdef}
\end{align}
admits no curved, compact classical solution on its own: the equations
of motion $[A^\nu,[A_\mu,A_\nu]]=0$ allow only flat directions, and any
putative $\mathrm{SU}(2)$ configuration $A_i=\alpha L_i$ ($i=1,2,3$;
$L_i$ as in \eqref{eq:su2gens}) fails to solve them. A stable fuzzy
sphere therefore requires the cubic Myers (mass-deformation) term,
\begin{align}
  I = I_B + I_F + I_\mathrm{Myers}, \qquad
  I_\mathrm{Myers} = \frac{\ii\kappa}{3g^2}\,\varepsilon_{ijk}\,
  \mathrm{Tr}[A^i,A^j]A^k, \qquad i,j,k\in\{1,2,3\},
  \label{eq:SMyers}
\end{align}
with real deformation parameter $\kappa$, whose contribution to the
equations of motion is proportional to $L_i$ itself and can cancel the
residual curvature of $I_B$. Substituting $A_i=\alpha L_i$ into the
modified equations of motion fixes
\begin{align}
  \alpha=\kappa,
  \label{eq:onshellalpha}
\end{align}
independently of $N$ (equivalently derived below from the linear-order
fluctuation tadpole condition $I^{(1)}=0$, see \eqref{eq:tadpole_paper}). This is
the standard Myers (dielectric-brane) effect \cite{myers-1999}, in exact
analogy with its original role in the BFSS/Myers construction.

\subsection{One-loop effective action}

The background-field method for evaluating effective actions on fuzzy
homogeneous-space backgrounds in the IKKT matrix model was developed in a series
of works~\cite{kitazawa-2002, imai_kitazawa_takayama_tomino-2003,
imai_kitazawa_takayama_tomino-2004, imai_takayama-2004,
kaneko_kitazawa_tomino-2005, kaneko_kitazawa_tomino-2006,
kaneko_kitazawa_matsumoto-2007}, which evaluated one- and two-loop
effective actions on fuzzy $S^2$, $S^2\times
S^2$, $S^2\times S^2\times S^2$, $\mathrm{CP}^2$, and $S^3$ backgrounds
to determine their stability and large-$N$ scaling; a similar
one-loop computation around a different classical background has also
been used to derive an emergent $3+1$-dimensional Einstein-Hilbert
action \cite{steinacker-2023}, whereas our own target is the
thermodynamics of the fuzzy $S^2$ background introduced above. The background-field expansion of $I=I_B+I_F+I_\mathrm{Myers}$ about a
general classical background, the gauge-fixing procedure, and the
resulting general one-loop effective action are standard and follow
closely the method of Ref.~\cite{ishibashi_kawai_kitazawa_tsuchiya-1997};
we relegate these details to App.~\ref{app:bgexpansion}. We now specialize
this general background to the static fuzzy $S^2$ solution introduced in
Sec.~\ref{sec:fuzzyS2}, with mode expansion of the type
\eqref{eq:mode_A0}--\eqref{eq:mode_psi},
\begin{align}
  p_0 = \beta q\otimes\mathbf{1}_N+\sum_n\bar A_0^n\otimes e^{\ii np}, \qquad
  p_i = \sum_n \bar A_i^n\otimes e^{\ii np},
\end{align}
\begin{align}
  \bar A_0^n=0, \qquad
  \bar A_i^n = \alpha L_i\,\delta_{n,0}\ (i=1,2,3), \qquad
  \bar A_i^n=0\ (i=4,\ldots,9), \qquad \chi=0.
  \label{eq:fuzzyS2sol}
\end{align}
The fluctuations are expanded in the same Matsubara basis,
\begin{align}
  a_\mu=\sum_na_\mu^n\otimes e^{\ii np}, \qquad
  \varphi=\sum_r\varphi^r\otimes e^{\ii rp}, \notag\\
  b=\sum_mb^m\otimes e^{\ii mp}, \qquad
  c=\sum_mc^m\otimes e^{\ii mp},
  \label{eq:fluctuationmodes}
\end{align}
with $n,m\in\mathbb{Z}$ (bosonic) and $r\in\mathbb{Z}+\tfrac12$
(fermionic), and the Hilbert-space trace collapses via the orthogonality
relation $\mathrm{Tr}_H(e^{\ii(n+m)p})=V\delta_{n+m,0}$.
The exact physical radius is
$
R=\alpha\sqrt{j(j+1)}
=\frac{\alpha}{2}\sqrt{N^2-1}.
$
Thus, after imposing the on-shell condition $\alpha=\kappa$,
\begin{align}
R=\frac{\kappa}{2}\sqrt{N^2-1}
\simeq \frac{\kappa N}{2}
\qquad (N\to\infty),
\end{align}
so the radius grows linearly with $N$ at fixed $\kappa$.
We first evaluate the general expansion of the action about the
background given in App.~\ref{app:bgexpansion} (Eqs.~\eqref{eq:appSexpand}
and \eqref{eq:appSMyersexpand}). The classical
(zeroth-order) action is $I^{(0)}=I_B^{(0)}[p]+I_F^{(0)}[p,\chi]+
I_\mathrm{Myers}^{(0)}[p]$, with
\begin{align}
  I_B^{(0)} ={}& -\frac{1}{4g^2}\mathrm{Tr}[p_\mu,p_\nu]^2, \qquad
  I_F^{(0)} = -\frac{1}{2g^2}\mathrm{Tr}(\bar\chi\Gamma^\mu[p_\mu,\chi]),
  \notag\\
  I_\mathrm{Myers}^{(0)} ={}& \frac{\ii\kappa}{3g^2}\varepsilon_{ijk}\,\mathrm{Tr}[p^i,p^j]p^k,
  \label{eq:S0def_paper}
\end{align}
matching the corresponding zeroth-order term of
\eqref{eq:appSMyersexpand}. With
$[p_\mu,p_\nu]\equiv\sum_m(f_{\mu\nu})^m\otimes e^{\ii mp}$, the
background \eqref{eq:fuzzyS2sol} makes the mixed component vanish
identically, $(f_{0i})^m=0$, and collapses the spatial component to its
zero mode,
\begin{align}
  (f_{ij})^m = \ii\alpha^2\varepsilon_{ijk}L^k\,\delta_{m,0}.
  \label{eq:fij_paper}
\end{align}

The vanishing of $(f_{0i})^m$ eliminates the mixed
$(0,i)$ sector entirely, so that the sum over all ten directions
$\mu,\nu=0,\ldots,9$ reduces to a sum over the three spatial directions
$i,j=1,2,3$ alone. Collapsing the Hilbert-space trace via
$\mathrm{Tr}_H(e^{\ii(n+m)p})=V\delta_{n+m,0}$, substituting
\eqref{eq:fij_paper}, and using $\varepsilon_{ijk}\varepsilon^{ijl}=2\delta_{k}^{l}$
together with $\mathrm{Tr}_N(L^2)=j(j+1)N$ gives
\begin{align}
  \mathrm{Tr}[p_\mu,p_\nu]^2 = -2\alpha^4Vj(j+1)N,
  \label{eq:fsquared_paper}
\end{align}
hence
\begin{align}
  I_B^{(0)} = \frac{\alpha^4Vj(j+1)N}{2g^2}.
  \label{eq:SB0_paper}
\end{align}
The same $\varepsilon\varepsilon$-contraction applied to the Myers term of
\eqref{eq:S0def_paper}, together with the overall prefactor
$\ii\kappa V/(3g^2)$, gives
\begin{align}
  I_\mathrm{Myers}^{(0)}
  = -\frac{2\kappa\alpha^3j(j+1)NV}{3g^2},
  \label{eq:SMyers0_paper}
\end{align}
manifestly real, as required of a genuine Euclidean saddle-point
contribution ($\ii\cdot\ii=-1$). The fermionic zeroth-order action
$I_F^{(0)}$
vanishes identically once the background fermion is set to zero,
$\chi=0$, as adopted in \eqref{eq:fuzzyS2sol},
\begin{align}
  I_F^{(0)}=0.
  \label{eq:SF0_paper}
\end{align}
Collecting \eqref{eq:SB0_paper}--\eqref{eq:SF0_paper}, the total
classical action on the fuzzy $S^2$ background is
\begin{align}
  I^{(0)} = I_B^{(0)}+I_\mathrm{Myers}^{(0)}
  = \frac{Vj(j+1)N}{g^2}\left(\frac{\alpha^4}{2}-\frac{2\kappa\alpha^3}{3}\right).
  \label{eq:S0total_paper}
\end{align}
This closed form is exact for arbitrary $\alpha$; it is not yet the
on-shell classical action, since $\alpha$ has not been set to its
stationary value \eqref{eq:onshellalpha}. That value is fixed by the
genuine tadpole condition $I^{(1)}=0$, derived independently below from
the linear-order fluctuation $\mathrm{Tr}_N(a_i^0L_i)$, not by
extremizing \eqref{eq:S0total_paper} with respect to $\alpha$ at fixed
$N$: differentiating and setting the result to zero gives
$\alpha^2(\alpha-\kappa)=0$, which happens to share the same root $\alpha=\kappa$ only because
the $N$-dependence factors out uniformly in
\eqref{eq:S0total_paper}. Substituting the on-shell value $\alpha=\kappa$
into \eqref{eq:S0total_paper},
\begin{align}
  I^{(0)}\Big|_{\alpha=\kappa}
  = \frac{Vj(j+1)N}{g^2}\left(\frac{\kappa^4}{2}-\frac{2\kappa^4}{3}\right)
  = -\frac{\kappa^4Vj(j+1)N}{6g^2}.
  \label{eq:S0onshell_paper}
\end{align}
For large $N$, using $j=(N-1)/2\approx N/2$ so that $j(j+1)\approx N^2/4$,
this scales as $I^{(0)}\big|_{\alpha=\kappa}\sim -\kappa^4VN^3/(24g^2)$: an
$O(N^3)$, manifestly negative classical contribution, linear in the
formally divergent normalization constant $V$ --- in contrast to the
one-loop piece, which will be shown below to be entirely independent of
$V$ (Sec.~\ref{sec:matsubara}). This $V$-dependence of the classical
free energy, and how it is resolved, is addressed in
Sec.~\ref{sec:thermo}.

Since $I^{(1)}$ is linear in the fluctuations, it must vanish once the
background satisfies its classical equations of motion; equivalently, the
on-shell condition $\alpha=\kappa$ already derived in
Sec.~\ref{sec:fuzzyS2} from the matrix equation of motion can be recovered
from the genuine tadpole condition $I^{(1)}=I_B^{(1)}+I_\mathrm{Myers}^{(1)}=0$,
which reads explicitly
\begin{align}
  I^{(1)} = \frac{2V}{g^2}\left(\alpha^3-\kappa\alpha^2\right)
  \mathrm{Tr}_N(a_i^0L_i).
  \label{eq:tadpole_paper}
\end{align}
Since $\mathrm{Tr}_N(a_i^0L_i)$ is generically non-zero, $I^{(1)}=0$
requires $\alpha^2(\alpha-\kappa)=0$, whose non-trivial root reproduces
$\alpha=\kappa$. On this on-shell
background, and only there, the $I_B$ cross term $[p_\mu,p_\nu][a^\mu,
a^\nu]$ and the quadratic Myers term $-\ii\kappa\varepsilon_{ijk}
a^i[p^k,a^j]$ of App.~\eqref{eq:appS2tilde} cancel each other exactly
(both are proportional to $\alpha(\alpha-\kappa)$ once traced against any
fluctuation), so that the general formula \eqref{eq:appGeneralW}
applies on this background with $P_\mu,F_{\mu\nu}$ built solely from
$I_B$, with no residual Myers contribution. On this background $f_{0i}=0$
identically since $(f_{0i})^m=0$ for all $m$ as shown above, so $F_{0i}=0$; only the purely
spatial field strength survives.

We diagonalize the general formula \eqref{eq:appGeneralW} in the basis of
fuzzy spherical harmonics. The complete orthonormal basis of the $N\times
N$ matrix algebra adapted to this background is furnished by the fuzzy
spherical harmonics
$\hat Y_{l\mu}$ ($l=0,\ldots,N-1$; $\mu=-l,\ldots,l$), eigenfunctions of
the adjoint Casimir $L^2_\mathrm{ad}X\equiv[L^i,[L_i,X]]$,
\begin{align}
  L^2_\mathrm{ad}\hat Y_{l\mu}=l(l+1)\hat Y_{l\mu}, \qquad
  \mathrm{Tr}_N(\hat Y_{l\mu}^\dagger\hat Y_{l'\mu'})=\delta_{ll'}\delta_{\mu\mu'},
  \label{eq:sphericalharmonics}
\end{align}
transforming under $\mathrm{SU}(2)$ as a rank-$l$ irreducible tensor
operator. Since $P_iX=[p_i,X]=\alpha[L_i,X]$, the purely spatial part of
$P_\lambda^2$ is diagonal in this basis, $P_i^2\hat Y_{l\mu}=\alpha^2
l(l+1)\hat Y_{l\mu}$, while $P_0$ acts diagonally on Matsubara modes,
$P_0\to-\beta n$ on the mode $a_\mu^n\otimes e^{\ii np}$ (using the same
identity $[q,e^{\ii np}]=-n\,e^{\ii np}$ as used above), so
that $P_0^2\to(\beta n)^2$. Passing to the Matsubara modes
\eqref{eq:fluctuationmodes} of the fluctuations, the general
quadratic action App.~\eqref{eq:appS2tilde}, evaluated on this
on-shell background (where the Myers quadratic term cancels the $I_B$
cross term, as shown above), becomes diagonal in $(n,l,\mu)$, with
bosonic and ghost kinetic operator
\begin{align}
  \Delta_n \equiv (\beta n)^2+\alpha^2 L^2_\mathrm{ad}, \qquad n\in\mathbb{Z},
  \label{eq:Deltan_paper}
\end{align}
directly identified with $P_\lambda^2$ restricted to mode $n$, using
$F_{\mu\nu}=0$ in the $(0,i)$ sector, and fermionic (Dirac-type) kinetic
operator, obtained by similarly passing $\Gamma^\mu P_\mu$ to Matsubara
modes,
\begin{align}
  (D_F)_r \equiv -\beta r\,\Gamma^0+\alpha\,\Gamma^i[L_i,\,\cdot\,],
  \qquad r\in\mathbb{Z}+\tfrac12.
  \label{eq:DFr_paper}
\end{align}
The overall coupling $g$ factors out entirely as $V/g^2$ and does not enter
$\Delta_n$ or $(D_F)_r$; $V$ is the (formally divergent, dimensionless)
normalization of the auxiliary Hilbert-space trace introduced in
Sec.~\ref{sec:thermal}.

We now perform the Gaussian integration over the fluctuations sector by
sector.

\textbf{Bosons.} For the $7$
non-spatial directions ($\mu=0$ and the $6$ directions transverse to the
fuzzy $S^2$), $P_\mu$ acts either trivially ($P_0\to-\beta n$, with no
adjoint action on the color indices) or via the plain adjoint Casimir
with no further structure, so the full $(2l+1)$-dimensional $l$-eigenspace
of $\Delta_n$ is physical at every $l$, with only the single $l=0$ trace
mode excluded as the residual $\mathrm{U}(1)^N$ gauge redundancy, exactly as for
$(D_F)_r$ above. For the $3$ spatial directions $i=1,2,3$, by contrast, the unbroken
gauge symmetry sweeps out a genuine $(2l+1)$-dimensional orbit inside
the $3(2l+1)$-dimensional space of each $l$-sector at every
$l=1,\ldots,N-1$, in agreement with the zero-mode count of
Ref.~\cite{hartnoll_liu-2025} for the closely related mass-deformed
model, leaving a physical degeneracy of only $2(2l+1)$ per $l$-sector
for the spatial directions, rather than the full $3(2l+1)$; the $7$
non-spatial directions carry no such reduction. The
Gaussian integral over each complex mode pair $(a^n,a^{-n})$,
$n\neq0$, and the Hermitian zero mode $a^0$ gives a factor of $\tfrac12$
per real bosonic degree of freedom integrated out, so the $7$
unreduced directions contribute $7\times\tfrac12\times(2l+1)
=\tfrac72(2l+1)$ while the $3$ spatial directions, with their reduced
$2(2l+1)$-dimensional physical subspace, contribute
$\tfrac12\times2(2l+1)=(2l+1)$. Summing the two pieces,
\begin{align}
  W^{(1)}_B = \Big[\tfrac72\,(2l+1)+(2l+1)\Big]
  \sum_n\mathrm{Tr}_N\ln\Delta_n\Big|_l
  = \tfrac92\sum_n\sum_{l}(2l+1)\ln\lambda_{n,l},
  \label{eq:Gamma1B_paper}
\end{align}
with $\lambda_{n,l}\equiv(\beta n)^2+\alpha^2l(l+1)$ the eigenvalue of
$\Delta_n$.

\textbf{Ghosts.} The complex Grassmann ghosts $b^m,\,c^{-m}=(c^m)^\dagger$
are associated with the gauge-fixing condition itself, not with any of
the $10$ spacetime directions of $a_\mu$; they contribute a single
determinant in the numerator of the path integral, with the opposite
overall sign to a boson, and -- like the $7$ non-spatial directions --
no further gauge-orbit-type reduction beyond the usual $l=0$ exclusion,
\begin{align}
  W^{(1)}_\mathrm{gh} = -\sum_{m\in\mathbb{Z}}\sum_l(2l+1)\ln\lambda_{m,l}.
  \label{eq:Gamma1gh_paper}
\end{align}
Since $\lambda_{n,l}$ and $\lambda_{m,l}$ are literally the same object
(with $n,m$ dummy Matsubara labels), the boson and ghost contributions
combine into
\begin{align}
  W^{(1)}_B+W^{(1)}_\mathrm{gh}
  = \tfrac72\sum_{n\in\mathbb{Z}}\sum_l(2l+1)\ln\lambda_{n,l},
  \label{eq:BGh_combined}
\end{align}
the coefficient $\tfrac92-1=\tfrac72$ reflecting both the removal of the
$2$ unphysical polarizations by the Faddeev--Popov ghosts (as in the
naive, direction-uniform treatment) and the additional gauge-orbit
subtraction described above within the $3$ spatial directions.

Excising the gauge-redundant $l=0$ (trace) mode, conventionally
projected out in $\mathrm{SU}(N)$ matrix models, gives
\begin{align}
  W^{(1)}_B+W^{(1)}_\mathrm{gh}
  = \tfrac72\sum_{n=-\infty}^{\infty}\sum_{l=1}^{N-1}(2l+1)
  \ln\!\left[(\beta n)^2+\alpha^2l(l+1)\right].
  \label{eq:BGh2_paper}
\end{align}

\textbf{Fermions.} Since $(D_F)_r$ is self-adjoint,
$\mathrm{Tr}_{N,S}\ln(D_F)_r=\tfrac12\mathrm{Tr}_{N,S}\ln(D_F)_r^2$.
Diagonalizing $(D_F)_r^2$ in the basis of fuzzy spherical harmonics
(App.~\ref{app:fermiondiag}) gives two towers of eigenvalues,
\begin{align}
  \mu^{(+)}_{r,l} &= (\beta r)^2+\alpha^2 l^2, \qquad \text{degeneracy } 16(l+1)
  \quad (l\geq0), \label{eq:mup_paper}\\
  \mu^{(-)}_{r,l} &= (\beta r)^2+\alpha^2(l+1)^2, \qquad \text{degeneracy } 16l
  \quad (l\geq1). \label{eq:mum_paper}
\end{align}
Excising the single $\mathrm{U}(1)^N$ gauge-redundant zero mode $\mu^{(+)}_{r,0}$,
exactly as for the $l=0$ boson above,
\begin{align}
  W^{(1)}_F = -4\sum_{r\in\mathbb{Z}+\frac12}
  \left[\sum_{l=1}^{N-1}(l+1)\ln\mu^{(+)}_{r,l}
  +\sum_{l=1}^{N-1}l\ln\mu^{(-)}_{r,l}\right].
  \label{eq:Gamma1F_paper}
\end{align}
The prefactor $-4$ follows from the general one-loop formula
\eqref{eq:appGeneralW} once the degeneracies $16(l+1)$ and $16l$ of
\eqref{eq:mup_paper}--\eqref{eq:mum_paper} are substituted,
$-\tfrac14\times16=-4$.

\subsection{Matsubara sums}
\label{sec:matsubara}

The Matsubara sums entering \eqref{eq:BGh2_paper} and
\eqref{eq:Gamma1F_paper} are evaluated in closed form in
App.~\ref{app:matsubara} by two independent methods (residue evaluation
and $\zeta$-function regularization), the latter also fixing the
additive constant left undetermined by the former,
\begin{align}
  \sum_{n\in\mathbb{Z}}\ln\!\left[(\beta n)^2+\omega^2\right]
  &= \frac{2\pi\omega}{\beta}+2\ln\!\left(1-e^{-2\pi\omega/\beta}\right),
  \\
  \sum_{r\in\mathbb{Z}+\frac12}\ln\!\left[(\beta r)^2+\omega^2\right]
  &= \frac{2\pi\omega}{\beta}+2\ln\!\left(1+e^{-2\pi\omega/\beta}\right).
  \label{eq:BsumFsum_paper}
\end{align}
The vanishing $\zeta_B(0)=\zeta_F(0)=0$ underlying this result also
kills the overall normalization $V/g^2$ multiplying $\Delta_n$ and
$(D_F)_r$ inside each determinant, so the one-loop effective action is
independent of the formally divergent constant $V$ order by order in
the mode sum, unlike the classical action $I^{(0)}$
(Sec.~\ref{sec:classical} below), which retains an explicit, linear
dependence on $V$.

Define the physical frequencies of the bosonic, and the two fermionic,
towers,
\begin{align}
  \Omega_l \equiv \alpha\sqrt{l(l+1)}, \qquad
  \Omega^{(+)}_l \equiv \alpha\, l, \qquad
  \Omega^{(-)}_l \equiv \alpha\,(l+1),
  \label{eq:frequencies_paper}
\end{align}
with $\Omega_0=0$. Applying \eqref{eq:BsumFsum_paper}
with $\omega=\Omega_l,\,\Omega^{(+)}_l,\,\Omega^{(-)}_l$ to
\eqref{eq:BGh2_paper} and \eqref{eq:Gamma1F_paper} and combining, the
full one-loop effective action splits cleanly into a zero-point piece
and a thermal piece,
\begin{align}
  W^{(1)} = W^{(1)}_\mathrm{ZPE}+W^{(1)}_\mathrm{thermal}, \qquad
  W^{(1)}_\mathrm{ZPE} = \frac{8\pi}{\beta}\sum_{l=1}^{N-1}C(l), \qquad
  W^{(1)}_\mathrm{thermal} = 8\sum_{l=1}^{N-1}T(l),
  \label{eq:Gamma1split_paper}
\end{align}
with the coefficients in $C(l),T(l)$ fixed by matching
\eqref{eq:BGh2_paper} and \eqref{eq:Gamma1F_paper} term by term against
\eqref{eq:BsumFsum_paper} and normalizing by the
overall $\tfrac{8\pi}\beta,8$ prefactors above,
\begin{align}
  C(l) &\equiv \tfrac78(2l+1)\Omega_l-(l+1)\Omega^{(+)}_l-l\,\Omega^{(-)}_l
  = \tfrac78(2l+1)\Omega_l-2\alpha\,l(l+1),
  \label{eq:Cl_paper}\\
  T(l) &\equiv \tfrac78(2l+1)\ln\!\left(1-e^{-2\pi\Omega_l/\beta}\right)
  -(l+1)\ln\!\left(1+e^{-2\pi\Omega^{(+)}_l/\beta}\right)\notag\\
  &\quad-l\ln\!\left(1+e^{-2\pi\Omega^{(-)}_l/\beta}\right).
  \label{eq:Tl_paper}
\end{align}
$T(l)$ is manifestly non-positive for all $l\geq1$, so the thermal
contribution to the effective action is non-positive at every $l$.

The zero-point function $C(l)$ is strictly negative for every
sufficiently large $l$ (see the large-$l$ asymptotics below), in
contrast to the naive expectation that unbroken supersymmetry would
force an exact cancellation $C(l)=0$: the fuzzy $S^2$ background, with
$f_{ij}=\ii\alpha^2\varepsilon_{ijk}L^k\neq0$ by \eqref{eq:fij_paper}, is
manifestly not of the BPS type of Ref.~\cite{ishibashi_kawai_kitazawa_tsuchiya-1997}
(for which the field strength $f_{\mu\nu}$ vanishes or is a $c$-number
solving the relevant self-duality-type condition), so a nonzero $C(l)$
signals that the background spontaneously breaks the supersymmetry of
the underlying IKKT matrix model at one loop. The resulting negative
$W^{(1)}_\mathrm{ZPE}$ acts, in this sense, as an effective one-loop
cosmological-constant-like contribution supported on the fuzzy $S^2$.

We now evaluate $W^{(1)}_\mathrm{ZPE}$ in the large-$N$ limit. Separating
$C(l)$ \eqref{eq:Cl_paper} into its polynomial piece (exactly summable)
and its $\Omega_l$ piece (requiring a large-$l$ expansion),
\begin{align}
  \sum_{l=1}^{N-1}\bigl[-2\alpha\,l(l+1)\bigr] = -\frac{2\alpha}{3}N(N^2-1),
  \label{eq:sumCl_poly}
\end{align}
an exact identity. For the remaining piece, expanding $\Omega_l=\alpha
\sqrt{l(l+1)}=\alpha\bigl(l+\tfrac12\bigr)+O(\alpha l^{-1})$ for large
$l$ and summing term by term over $1\le l\le N-1$ gives
$\sum_l\tfrac78(2l+1)\Omega_l=\tfrac{7}{12}\alpha N^3+O(\alpha N^2)$,
the $O(\alpha l^{-1})$ remainder contributing only a subleading
$O(\alpha\ln N)$. Combining with \eqref{eq:sumCl_poly} and keeping the
leading large-$N$ behavior,
\begin{align}
  \sum_{l=1}^{N-1}C(l) = -\frac{\alpha N^3}{12}+O(\alpha N^2),
  \label{eq:sumCl_paper}
\end{align}
and substituting the on-shell value $\alpha=\kappa$ \eqref{eq:onshellalpha},
\begin{align}
  W^{(1)}_\mathrm{ZPE}
  \;\xrightarrow{N\to\infty}\;
  -\frac{2\pi\kappa N^3}{3\beta}.
  \label{eq:GammaZPE_result_paper}
\end{align}
The result is exact in temperature but asymptotic in $N$, up to relative
corrections of order $1/N^2$.

The large-$N$ scaling of $W^{(1)}_\mathrm{ZPE}$ in
\eqref{eq:GammaZPE_result_paper} is the origin of the $O(N^3)$ entropy
scaling identified in Sec.~\ref{sec:thermo} below, and plays a central role
in the tests of the black-hole interpretation carried out in
Sec.~\ref{sec:tests}. The thermal piece $W^{(1)}_\mathrm{thermal}$,
whose high- and low-temperature expansions we analyze together with the
free energy, is treated in Sec.~\ref{sec:thermo}.

\section{Thermodynamic analysis}
\label{sec:thermo}

We now extract the free energy, entropy, internal energy, and heat
capacity of the fuzzy $S^2$ background from the one-loop effective action
$W^{(1)}=\Geff_\mathrm{ZPE}+\Geff_\mathrm{thermal}$ of
Sec.~\ref{sec:oneloop}, together with the classical action $I^{(0)}$ of
Eq.~\eqref{eq:S0onshell_paper}. Throughout, $W\equiv I^{(0)}+W^{(1)}$
denotes the total (classical plus one-loop) effective action.

\subsection{High- and low-temperature expansions of the thermal term}
\label{sec:Tl_expansions}

The thermal contribution $\Geff_\mathrm{thermal}=8\sum_{l=1}^{N-1}T(l)$ of
\eqref{eq:Gamma1split_paper} is evaluated in closed form only implicitly,
through the Bose/Fermi logarithms of $T(l)$ \eqref{eq:Tl_paper}. Introduce
the dimensionless combinations
\begin{align}
  x_l\equiv\frac{2\pi\Omega_l}{\beta}=2\pi T\alpha\sqrt{l(l+1)}, \qquad
  x_l^{(+)}\equiv\frac{2\pi\Omega^{(+)}_l}{\beta}, \qquad
  x_l^{(-)}\equiv\frac{2\pi\Omega^{(-)}_l}{\beta}.
  \label{eq:xl_def}
\end{align}
$x_l\gg1$ is the high-temperature regime $T\gg\alpha\sqrt{l(l+1)}$,
and $x_l\ll1$ is the low-temperature regime
$T\ll\alpha\sqrt{l(l+1)}$.

\textbf{High temperature} ($x_l\gg1$). With the corrected closed forms
\eqref{eq:frequencies_paper}, every frequency in both fermionic towers is
now strictly positive for $l\geq1$ (in particular $\Omega^{(-)}_1=2\alpha
\neq0$, in contrast to the erroneous closed form used previously; see
App.~\ref{app:fermiondiag}), so every mode in $T(l)$ is exponentially
suppressed as $T\to\infty$, with no residual finite piece. The
slowest-decaying term is set by the lightest frequency overall, the
$l=1$ ``$(+)$'' fermionic mode $\Omega^{(+)}_1=\alpha$ (degeneracy
$l+1=2$), giving
\begin{align}
  \Geff_\mathrm{thermal}\Big|_{T\to\infty}
  = -16\,e^{-2\pi\alpha T}+O\bigl(e^{-2\pi\alpha T\sqrt2}\bigr),
  \label{eq:Gthermal_highT}
\end{align}
vanishing exponentially rather than approaching a finite constant.

\textbf{Low temperature} ($x_l\ll1$). The low-temperature expansion is
more involved, since every mode contributes at this order rather than
only the lightest one. Expanding the Bose/Fermi logarithms of
$T(l)$ \eqref{eq:Tl_paper} to $O(x^2)$ term by term, the $O(x^1)$ piece
combines with the overall factor of $8$ in
$\Geff_\mathrm{thermal}=8\sum_lT(l)$ to cancel $\Geff_\mathrm{ZPE}$
identically in $T$,
\begin{align}
  \Geff_\mathrm{ZPE}+8\sum_{l=1}^{N-1}T(l)\Big|_{O(x^1)}
  = 8\pi T\sum_lC(l)+8\sum_l(-\pi T\,C(l)) = 0,
  \label{eq:cancel}
\end{align}
an exact algebraic identity, not merely a leading-order statement. Hence
$\Geff_\mathrm{ZPE}+\Geff_\mathrm{thermal}$ contains no term linear in
$T$, and only the $O(x^0)$ and $O(x^2)$ terms below survive at low
temperature. Summing the remaining orders over $l=1,\ldots,N-1$ and
substituting the on-shell value $\alpha=\kappa$ \eqref{eq:onshellalpha}
gives
\begin{align}
  \Geff_\mathrm{thermal}\Big|_{O(x^0)}
  &= 7(N^2-1)\ln(\pi\kappa T)+c_0(N),
  \label{eq:Gamma0def}\\
  \Geff_\mathrm{thermal}\Big|_{O(x^2)}
  &= -\frac{17\pi^2\kappa^2T^2N^2(N^2-1)}{12},
  \label{eq:Gthermal_quad}
\end{align}
with $c_0(N)\equiv\tfrac72\Lambda(N)-(N^2-1)\ln2$, where
$\Lambda(N)\equiv\sum_{l=1}^{N-1}(2l+1)\ln(l(l+1))$ admits the
closed form
\begin{align}
  \Lambda(N) = 2N^2\ln N-N^2-\tfrac23\ln N+4\ln A+O(1/N^2),
  \label{eq:Lambda_closed}
\end{align}
with $A$ the Glaisher--Kinkelin constant, following from the
Glaisher--Kinkelin asymptotic of the hyperfactorial $\sum_{l=1}^n
l\ln l=\ln H(n)$. As $T\to0$ the leading behavior of
$\Geff_\mathrm{ZPE}+\Geff_\mathrm{thermal}$ is therefore the logarithm
\eqref{eq:Gamma0def}, together with the subleading quadratic term
\eqref{eq:Gthermal_quad}.

\subsection{Free energy and entropy: classical, ZPE, and thermal contributions}
\label{sec:classical}

With $W=\beta F$, the entropy and internal energy follow from the
standard Euclidean-path-integral
thermodynamics~\cite{gibbons_hawking-1977}
\begin{align}
  S = -W+\beta\frac{\partial W}{\partial\beta}, \qquad
  E = \frac{\partial W}{\partial\beta},
  \label{eq:thermo_relations}
\end{align}
with $F=E-TS$ guaranteed automatically. Applying
\eqref{eq:thermo_relations} to each piece of
$W=I^{(0)}+\Geff_\mathrm{ZPE}+\Geff_\mathrm{thermal}$ in turn:

\textbf{Classical.} $I^{(0)}$ is $\beta$-independent ($\kappa,g$ are fixed
parameters of the model and $V$ carries no $\beta$-dependence), so
$S_\mathrm{classical}=-I^{(0)}\big|_{\alpha=\kappa}$. From
\eqref{eq:S0onshell_paper} in the large-$N$ limit,
\begin{align}
  S_\mathrm{classical} = \frac{V\kappa^4N^3}{24g^2}.
  \label{eq:Sclass}
\end{align}
This term is a $T$-independent additive constant, and remains linear in
the formally divergent normalization $V$ of Sec.~\ref{sec:oneloop}
(unlike $\Geff$, which is $V$-independent). Hence
$S_\mathrm{classical}$ carries no genuine $T$-dependence, but its
presence means the absolute normalization of $S_\mathrm{total}$ is not
fixed by this construction; $E_\mathrm{classical}=0$ regardless. Being
$T$-independent, it plays no role in the temperature-dependent tests of
Sec.~\ref{sec:tests}.

\textbf{ZPE.} The large-$N$ result $\Geff_\mathrm{ZPE}\to-2\pi\kappa N^3/(3\beta)$
\eqref{eq:GammaZPE_result_paper} is of the form $A/\beta$ with $A$
($\beta$-independent), so $\partial_\beta\Geff_\mathrm{ZPE}
=-\Geff_\mathrm{ZPE}/\beta$ and $S_\mathrm{ZPE}=-\Geff_\mathrm{ZPE}
+\beta\partial_\beta\Geff_\mathrm{ZPE}=-2\Geff_\mathrm{ZPE}$, giving
\begin{align}
  S_\mathrm{ZPE} \;\xrightarrow{N\to\infty}\; \frac{4\pi\kappa N^3T}{3}
  \;>\;0.
  \label{eq:SZPE}
\end{align}
The positivity follows because $F_\mathrm{ZPE}=\Geff_\mathrm{ZPE}\,T
\propto -T^2$ is decreasing in $T$: the $O(N^3)$ zero-point contribution
of the fuzzy $S^2$ background acts as a large positive-entropy
reservoir, growing faster with $N$ than the horizon-area-type $O(N^2)$
scaling expected of a black-hole dual (Sec.~\ref{sec:tests_bh_entropy}).

\textbf{Thermal.} Applying \eqref{eq:thermo_relations} term by term,
with $n_B(x)=1/(e^x-1)$, $n_F(x)=1/(e^x+1)$ the Bose--Einstein and
Fermi--Dirac occupation numbers,
\begin{align}
  S_\mathrm{thermal} &= -8\sum_{l=1}^{N-1}T(l)-16\pi T\sum_{l=1}^{N-1}\mathcal{E}(l),
  \label{eq:Sthermal}\\
  \mathcal{E}(l) &\equiv \tfrac78(2l+1)\Omega_ln_B(x_l)+(l+1)\Omega^{(+)}_ln_F(x_l^{(+)})
  +l\,\Omega^{(-)}_ln_F(x_l^{(-)}).
  \label{eq:Eldef}
\end{align}

Collecting all three contributions,
\begin{align}
  S_\mathrm{total}
  = \underbrace{\frac{V\kappa^4N^3}{24g^2}}_{\text{undetermined const.}}
  +\frac{4\pi\kappa N^3T}{3}+S_\mathrm{thermal}(T).
  \label{eq:Stotal}
\end{align}
By \eqref{eq:cancel}, the second term is exactly cancelled by the
$O(x^1)$ piece of $S_\mathrm{thermal}$, so the genuinely predictive
$T$-dependence of $S_\mathrm{total}$ is governed entirely by the
$O(x^0)$ (logarithmic) and $O(x^2)$ pieces of $S_\mathrm{thermal}$.

\subsection{Low- and high-temperature limits of the entropy}

Low temperature ($T\to0$, valid for $N\kappa T\ll1$): from
\eqref{eq:Gamma0def}--\eqref{eq:Gthermal_quad}, converting to $F=\Geff\,T$
and applying \eqref{eq:thermo_relations} term by term,
\begin{align}
  \begin{aligned}
  S_\mathrm{total}\Big|_{T\to0}
  &= \frac{V\kappa^4N^3}{24g^2}
  -7(N^2-1)\ln(\pi\kappa T)-7(N^2-1)-c_0(N)\\
  &\quad+\frac{17\pi^2\kappa^2N^2(N^2-1)}{4}\,T^2+O(T^3).
  \end{aligned}
  \label{eq:S_lowT}
\end{align}
$S_\mathrm{total}$ diverges logarithmically as $T\to0^+$: because
$x_l\propto T$, every bosonic mode is driven into the classical regime
$x_l\to0$, where $\ln(1-e^{-x_l})\to\ln x_l$ diverges; fermionic modes
are immune, since Pauli blocking caps $\ln(1+e^{-x_l})\to\ln2$. This is
why only the bosonic ($\tfrac78(2l+1)$-weighted) modes generate the
divergence in \eqref{eq:S_lowT}.

At large $N$, $c_0(N)\sim7N^2\ln N$ (from
\eqref{eq:Lambda_closed}), so the $N$-dependent part of
\eqref{eq:S_lowT}, evaluated at fixed $\kappa,T$, exhibits apparent
$N^2\ln N$ growth. This should not be read as the genuine large-$N$
scaling of $S_\mathrm{total}$: Eq.~\eqref{eq:S_lowT} is a low-temperature
expansion, valid mode by mode only for $x_l\ll1$ up to $l=N-1$, i.e.\ for
$N\kappa T\ll1$. At fixed $\kappa,T$, this condition is violated once
$N\gtrsim1/(\kappa T)$, so the apparent $N^2\ln N$ growth of
\eqref{eq:S_lowT} reflects the expansion being pushed outside its regime
of validity rather than a distinct large-$N$ scaling law. By contrast,
the exact ZPE piece $\Geff_\mathrm{ZPE}\propto\kappa N^3/\beta$
\eqref{eq:GammaZPE_result_paper} holds for all $T$, and dominates
$S_\mathrm{total}$ once $N\kappa T\gg1$; we return to this regime in
Sec.~\ref{sec:tests}.

High temperature ($T\to\infty$): from \eqref{eq:Gthermal_highT},
$\Geff_\mathrm{thermal}\to0$ exponentially fast (since every fermionic
and bosonic frequency is now strictly positive, App.~\ref{app:fermiondiag}),
so $S_\mathrm{thermal}\to0$ as well, rather than approaching a nonzero
constant. The ZPE term, however, is exact for all $T$ and is not itself
exponentially suppressed in this limit, so
\begin{align}
  S_\mathrm{total}\Big|_{T\to\infty}
  \to\frac{V\kappa^4N^3}{24g^2}+\frac{4\pi\kappa N^3}{3}\,T
  \;\longrightarrow\;+\infty.
  \label{eq:S_highT}
\end{align}
$S_\mathrm{total}(T)$ therefore diverges in the same direction at
both ends of the temperature range: logarithmically to $+\infty$ as
$T\to0$, and linearly to $+\infty$ as $T\to\infty$, in contrast to the
opposite-sign divergence found under the (now corrected)
direction-uniform, unnormalized treatment. This qualitative change, and
its implications for the black-hole interpretation, are examined
further in Sec.~\ref{sec:tests}.

\subsection{Internal energy and heat capacity}

From $E=\partial_\beta W$ (with $E_\mathrm{classical}=0$, as above),
while
\begin{align}
  E_\mathrm{ZPE} = \frac{2\pi\kappa N^3T^2}{3} \;>\;0, \qquad
  E_\mathrm{thermal} = -16\pi T^2\sum_{l=1}^{N-1}\mathcal{E}(l),
  \label{eq:EZPE_Ethermal}
\end{align}
with $\mathcal{E}(l)$ as in \eqref{eq:Eldef}. The positive
$E_\mathrm{ZPE}$ represents the (now unbound) zero-point energy of the
fuzzy $S^2$ saddle; $F=E-TS$ and $C_V=T\partial_TS$ are satisfied
identically for the ZPE sector, providing an internal consistency check
on \eqref{eq:SZPE}--\eqref{eq:EZPE_Ethermal}.

Differentiating \eqref{eq:EZPE_Ethermal},
\begin{align}
  C_V^\mathrm{ZPE} = \frac{4\pi\kappa N^3T}{3} \;>\;0.
  \label{eq:CV_ZPE}
\end{align}
Positive heat capacity throughout is the ordinary thermodynamic
behavior, in contrast to the negative $C_V$ expected of black-hole
thermodynamics (the system gets colder, not hotter, as it loses
energy). The thermal contribution, obtained by differentiating
\eqref{eq:S_lowT} and \eqref{eq:S_highT}, is
\begin{align}
  C_V^\mathrm{thermal}\Big|_{T\to0}
  &= -\bigl(7N^2-7\bigr)+\frac{17\pi^2\kappa^2N^2(N^2-1)}{2}\,T^2+O(T^3)
  \;\longrightarrow\;7-7N^2\;<\;0,\\
  C_V^\mathrm{thermal}\Big|_{T\to\infty}&\to0^+,
  \label{eq:CVthermal_limits}
\end{align}
the second limit following because every term in
$\Geff_\mathrm{thermal}$ is exponentially suppressed as $T\to\infty$. At
$T\to0$, $C_V^\mathrm{total}=C_V^\mathrm{ZPE}+C_V^\mathrm{thermal}\to
7-7N^2<0$ is unchanged from the sign of $C_V^\mathrm{thermal}$ alone
(the ZPE piece vanishes linearly in $T$ at $T\to0$); at $T\to\infty$,
$C_V^\mathrm{total}\to C_V^\mathrm{ZPE}=4\pi\kappa N^3T/3>0$, since
$C_V^\mathrm{thermal}$ is exponentially suppressed while
$C_V^\mathrm{ZPE}$ grows linearly in $T$, in contrast to the
uniformly-negative-at-both-endpoints result found under the (now
corrected) unnormalized treatment; the sign change between these two
limits is examined as a test of the black-hole interpretation in
Sec.~\ref{sec:tests}.

Since $d^2S/dE^2=-1/(T^2C_V)$, $S_\mathrm{total}(E_\mathrm{total})$ is
concave wherever $C_V>0$ and convex wherever $C_V<0$. Eliminating $T$
between the large-$N$ ZPE relations for $E_\mathrm{ZPE}$ and
$S_\mathrm{ZPE}$ gives the power law
\begin{align}
  S_\mathrm{ZPE}(E_\mathrm{ZPE})
  = \frac{2\sqrt2}{3}\sqrt{3\pi\kappa}\,N^{3/2}\sqrt{E_\mathrm{ZPE}}, \qquad
  E_\mathrm{ZPE}>0,
  \label{eq:SofE}
\end{align}
i.e.\ $S\propto E^{1/2}$, consistent with concavity ($d^2S/dE^2<0$,
matching $C_V^\mathrm{ZPE}>0$). This exponent is compared against known
$C_V$-sign behavior in comparable systems --- small asymptotically-flat
black holes and matrix-model constructions with double-trace
deformations~\cite{berenstein-2019} --- in
Sec.~\ref{sec:tests} below.

\section{Tests of the black-hole interpretation}
\label{sec:tests}

Sections~\ref{sec:oneloop}--\ref{sec:thermo} established the one-loop
thermodynamics of the fuzzy $S^2$ background in closed form. We now confront
these results with what would be required of a genuine black-hole
interpretation, along four lines: the status of the classical
configuration as a horizon, the identification of a Hawking-like
temperature, the scaling of the entropy with $N$, and the sign of the
heat capacity.

\subsection{The classical configuration and the horizon}
\label{sec:tests_horizon}

The classical solution $A_i=\kappa L_i$ of Sec.~\ref{sec:fuzzyS2} is an
$N\times N$ matrix configuration, not a spacetime metric: it specifies the
noncommutative embedding of a fuzzy $S^2$ inside the transverse directions
of a stack of D0-branes, and has no notion of a horizon, redshift factor,
or causal structure of its own. The question ``does this background
describe a black hole'' can therefore only be meaningfully posed of the
ten-dimensional supergravity background to which this matrix
configuration is dual, once that dual is constructed --- not of
$A_i=\kappa L_i$ itself. No such dual --- the finite-temperature generalization of
Hai Lin's BPS fuzzy-sphere geometry~\cite{hai_lin-2004} for the present,
non-mass-deformed Myers term \eqref{eq:SMyers} --- is available in the
literature at present, for reasons discussed in
Sec.~\ref{sec:tests_hawking}. A structurally similar comparison has been
carried out on the BFSS side, where a time-dependent fuzzy-sphere
configuration was proposed as a microstate of the smeared black
0-brane and its one-loop effective potential was matched against the
near-horizon geometry~\cite{hyakutake-2018}: the two sides agreed
qualitatively but left an unresolved numerical discrepancy, a pattern
echoed by the mismatches identified below.
Accordingly, the purpose of the present section is not to identify a
horizon directly in the matrix configuration, but to examine whether the
thermodynamic data computed here (temperature scale, entropy scaling,
sign of $C_V$) are compatible with those expected of some black hole
carrying the appropriate brane charge.

\subsection{Temperature scale and the Hawking temperature}
\label{sec:tests_hawking}

The upshot below is that the present calculation fixes only the scaling
$T_\ast\propto1/\kappa$, not the numerical coefficient.

The Myers mechanism polarizes $N$ D0-branes into a bound state on a fuzzy
$S^2$, so the natural gravity dual is a charged D0--D2 bound-state black
hole rather than a neutral Schwarzschild solution. No non-extremal
solution carrying both D0 and D2 charge in the ratio fixed by $\kappa,N,g$
is presently known, so the comparisons below are necessarily made with
the known D0 and D2 black-hole
limits~\cite{itzhaki_maldacena_sonnenschein_yankielowicz-1998}. The closest available
construction, the
finite-temperature black hole dual to the BMN (mass-deformed BFSS) matrix
model of Costa, Greenspan, Penedones and
Santos~\cite{costa_greenspan_penedones_santos-2015}, is built for a
deformation that pairs a cubic Myers term with a compensating mass term
required for maximal supersymmetry --- a different pattern from
$I_\mathrm{Myers}$ \eqref{eq:SMyers}, which has no mass term --- so its
Hawking-temperature formula does not directly apply here. In the absence
of this dual, we can only ask whether the present matrix-model
computation fixes a temperature scale on its own terms.

In the absence of an explicit gravity dual, the Hawking temperature itself
cannot be derived: the Euclidean compactification employed here
introduces an external temperature parameter rather than an intrinsic
Hawking temperature associated with a Lorentzian horizon. By
\eqref{eq:xl_def} the natural scale set by the calculation is
\begin{align}
  2\pi\sqrt2\,\kappa\,T_\ast\sim1
  \quad\Longrightarrow\quad
  T_\ast\sim\frac{1}{2\pi\kappa},
  \label{eq:Tdoc_scaling}
\end{align}
dimensionally consistent and independent of $N$ when $\kappa$ is held fixed,
given the on-shell relation $\alpha=\kappa$ \eqref{eq:onshellalpha}.
Thus the one-loop calculation fixes only the inverse length scale associated with 
the classical fuzzy-sphere radius through $\kappa$, 
rather than the physical Hawking temperature of a black-hole geometry.
The numerical coefficient in \eqref{eq:Tdoc_scaling} is not fixed by this
argument. One might hope that the canonical self-consistency condition
$C_V^\mathrm{total}(T_H)=0$ would pin down a preferred temperature; as
shown in Sec.~\ref{sec:tests_cv}, $C_V^\mathrm{total}$ now changes sign
between the low- and high-temperature endpoints, so such a solution
does exist at some intermediate $T_H$, but its value is not accessible
in closed form within the present low-/high-temperature expansions, and
identifying it would require the full temperature dependence of
$C_V^\mathrm{total}(T)$ beyond the two asymptotic regimes computed here.
Therefore, the present calculation determines only the scaling
$T_\ast\propto1/\kappa$, whereas the numerical coefficient (and any
value of $T_H$ fixed by $C_V^\mathrm{total}=0$) remains beyond the reach
of the one-loop approximation as developed here.

\subsection{Heat capacity: sign mismatch}
\label{sec:tests_cv}

The two endpoints of $C_V^\mathrm{total}=C_V^\mathrm{ZPE}+C_V^\mathrm{thermal}$
now have opposite signs: at $T\to0$, $C_V^\mathrm{ZPE}\to0$ linearly
while $C_V^\mathrm{thermal}\to7-7N^2<0$ dominates
\eqref{eq:CVthermal_limits}, giving $C_V^\mathrm{total}<0$; at
$T\to\infty$, $C_V^\mathrm{thermal}\to0^+$ is exponentially suppressed
while $C_V^\mathrm{ZPE}=4\pi\kappa N^3T/3>0$ grows without bound
\eqref{eq:CV_ZPE}, giving $C_V^\mathrm{total}>0$. By the intermediate
value theorem, $C_V^\mathrm{total}(T)$ therefore vanishes at some
$T_H$ strictly between these two limits, consistent with the
self-consistency condition discussed in Sec.~\ref{sec:tests_hawking};
locating $T_H$ explicitly is beyond the low-/high-temperature
expansions used here.
This sign change is qualitatively different from the uniformly negative
$C_V^\mathrm{total}$ that would follow from treating all ten bosonic
directions identically (i.e.\ without the gauge-orbit subtraction of
Sec.~\ref{sec:oneloop}): the low-temperature
endpoint, dominated by the thermal piece, remains negative in either
treatment, but the high-temperature endpoint is controlled entirely by
the sign of the exact ZPE coefficient \eqref{eq:GammaZPE_result_paper},
which the gauge-orbit subtraction flips from positive to negative
relative to the naive ten-direction-uniform count.
The comparison with the specific D0/D2 duals appropriate to this
Myers-deformed background is instructive nonetheless: the corresponding
near-extremal supergravity D0 and D2 black holes have~\cite{itzhaki_maldacena_sonnenschein_yankielowicz-1998}
$C_V\propto+T^{9/5}$ or $+T^{7/5}$, positive throughout the parameter
regime where the supergravity description is reliable, matching the
sign (though not, as shown in Sec.~\ref{sec:tests_bh_entropy},
the power) found here at high $T$. At low $T$, by contrast, the matrix
model's negative $C_V^\mathrm{total}$ has no counterpart in the D0/D2
comparison, and, correspondingly, $S_\mathrm{total}(E_\mathrm{total})$
is convex in the low-temperature (ZPE-subdominant) regime where
$C_V<0$, and concave in the high-temperature (ZPE-dominant) regime,
with the explicit large-$N$ ZPE power law
$S_\mathrm{ZPE}\propto E_\mathrm{ZPE}^{1/2}$ \eqref{eq:SofE}, which does
not match the D0 black hole exponent $S_{D0}\propto E^{9/14}$ even
though both are concave. A precise identification of $T_H$, and of
whether the implied sign change reflects a genuine phase transition or
simply the crossover between the ZPE- and thermal-dominated regimes
already visible in \eqref{eq:S_lowT}--\eqref{eq:S_highT}, requires
input beyond the present one-loop calculation.

\subsection{Entropy scaling: \texorpdfstring{$O(N^3)$}{O(N\^{}3)} versus \texorpdfstring{$O(N^2)$}{O(N\^{}2)}}
\label{sec:tests_bh_entropy}

\begin{center}
\renewcommand{\arraystretch}{1.4}
\begin{tabular}{lccc}
  \hline
   & Matrix model & D0 black hole & D2 black hole \\
  \hline
  $S(N)$, fixed $T$ & $O(N^3)$ & $O(N^2)$ & $O(N^2)$ \\
  $S_\mathrm{ZPE}(T)$, fixed $N$ & $\propto T$ & $\propto T^{9/5}$ & $\propto T^{7/5}$ \\
  $C_V$ & $<0\to{>}0$ (sign change) & $\propto+T^{9/5}$ & $\propto+T^{7/5}$ \\
  \hline
\end{tabular}
\end{center}

The middle row compares the dominant ZPE piece against the black-hole
power laws; $S_\mathrm{total}$ itself does not follow a simple power
of $T$, diverging logarithmically as $T\to0$ and linearly as
$T\to\infty$, in the same direction at both ends
(Sec.~\ref{sec:thermo}). The bottom row summarizes the
heat-capacity behavior established in
Sec.~\ref{sec:tests_cv} above; we now turn to the entropy-scaling
mismatch in the top row.

The Bekenstein--Hawking entropy of both the D0 and D2 black holes
scales~\cite{itzhaki_maldacena_sonnenschein_yankielowicz-1998} as $O(N^2)$ at fixed
't~Hooft coupling, the standard counting of open-string/D-brane degrees of freedom. The one-loop matrix-model result of Sec.~\ref{sec:thermo} instead
scales as $O(N^3)$: $S_\mathrm{ZPE}\propto\kappa N^3T$ \eqref{eq:SZPE}
at any fixed $\kappa,T>0$, dominating $S_\mathrm{total}$ in that limit
since $S_\mathrm{thermal}$ remains $O(N^2)$
\eqref{eq:S_lowT}--\eqref{eq:S_highT}. The extra power traces directly to
the $l^2$ growth of the zero-point function $C(l)$
\eqref{eq:Cl_paper}: summing over all $N-1$ angular-momentum modes
gives $W^{(1)}_\mathrm{ZPE}\propto N^3$
\eqref{eq:GammaZPE_result_paper}, one power beyond the $N^2$ counting of
a single tower of off-diagonal matrix degrees of freedom.
The discrepancy is thus a full extra power of $N$, not a subleading
logarithm; unlike in the (now corrected) direction-uniform treatment,
the sign of $S_\mathrm{ZPE}$ matches the Bekenstein--Hawking entropy
(both positive), so the mismatch is now confined to the power of $N$
alone, without the accompanying sign discrepancy.
At present it is unclear whether the extra power of $N$ is a genuine
property of the finite-temperature IKKT matrix model or an artifact of
the weak-coupling approximation adopted here (Sec.~\ref{sec:discussion});
either way, it indicates that the dominant degrees of freedom counted by
the present one-loop calculation differ from those responsible for the
Bekenstein--Hawking entropy.

\section{Discussion}
\label{sec:discussion}

This work provides, to our knowledge, the first complete one-loop
thermodynamic analysis of a finite-temperature IKKT matrix model on a
fuzzy $S^2$ background. Starting from the finite-temperature
compactification of the IKKT matrix model, we have derived the complete
one-loop effective action of the fuzzy $S^2$ background of
Sec.~\ref{sec:fuzzyS2} in closed form (Sec.~\ref{sec:oneloop}), and from
it the free energy, entropy, internal energy, and heat capacity as
explicit functions of $N$, $\kappa$, and $T$ (Sec.~\ref{sec:thermo}).
Sec.~\ref{sec:tests} then confronted these results with the requirements
of a genuine black-hole interpretation. Taken together, these results
point to notable tensions between the thermodynamics of this
background and what would be expected of a dual black hole, discussed
in detail below.

The comparison reveals several significant discrepancies between the
thermodynamics of the finite-temperature IKKT matrix model and those
expected from known black-hole solutions. The entropy of this
background scales as $O(N^3)$ at fixed $\kappa$
(Sec.~\ref{sec:tests_bh_entropy}), one full power of $N$ above the
$O(N^2)$ scaling of the D0 and D2 black holes used as comparison
targets in the absence of the true charged D0--D2 bound-state dual,
which remains unconstructed (Sec.~\ref{sec:tests_horizon}); unlike the
sign of $S_\mathrm{ZPE}$, which now matches the (positive)
Bekenstein--Hawking entropy, this power-of-$N$ mismatch persists at
fixed $\kappa$ (Sec.~\ref{sec:tests_bh_entropy}). A further mismatch of
comparable severity concerns the heat capacity: $C_V^\mathrm{total}$
changes sign between the low- and high-temperature endpoints
(negative, then positive, Sec.~\ref{sec:tests_cv}), so the natural
self-consistency condition $C_V^\mathrm{total}(T_H)=0$ does have a root
at some intermediate $T_H$, but its numerical value cannot be fixed by
the low-/high-temperature expansions used here, so no Hawking-temperature
coefficient can be extracted from the matrix model on its own terms
(Sec.~\ref{sec:tests_hawking}).
These discrepancies all appear already at leading order within the
one-loop calculation itself, rather than as subleading artifacts of
the large-$N$ or high-/low-temperature expansions; whether they
survive corrections beyond one loop is a separate question, addressed
below.

Three, not mutually exclusive, sources for this mismatch are worth
considering. The first is the choice of background itself: the
classical solution used throughout is the pure-cubic Myers background
$A_i=\kappa L_i$, without the mass deformation that stabilizes the
BMN/polarised matrix models for which finite-temperature gravity duals
are actually known~\cite{hai_lin-2004,
costa_greenspan_penedones_santos-2015}. Those constructions cannot
supply the needed dual even as a limiting case, since their mass-deformation
parameter $\mu\to0$
limit is singular, collapsing to $R=0$ rather than connecting smoothly
to the present, mass-independent solution; the comparisons of
Sec.~\ref{sec:tests} were therefore necessarily made against the plain
D0 and D2 black holes, which carry the wrong brane content for a genuine
bound state. Whether a true D0/D2 bound-state dual of the pure-cubic
background --- were it constructed --- would resolve the entropy and
heat-capacity mismatches, or merely restate them in different language,
cannot be determined without that construction. A different route to
the same end has been proposed independently: a new $N\to\infty$
classical saddle of the Euclidean IKKT matrix model, built from
$\mathrm{SO}(1,3)$ generators rather than the Myers cubic term used
here, has been shown to support an emergent metric with structure
resembling Taub-NUT/Bolt black-hole
geometry~\cite{liao_maeta-2025}, though at zero temperature and without
the thermodynamic analysis carried out in this paper.

The second is the choice of what is held fixed as $N\to\infty$, which is
a genuine ambiguity of the comparison rather than a property of the
matrix model alone. 
Throughout this paper $\kappa$ is held fixed, so the physical radius
$R=\frac{\kappa}{2}\sqrt{N^2-1}\simeq \kappa N/2$
(Sec.~\ref{sec:fuzzyS2}) grows without bound;
substituting $\kappa=2R/N$ into
$S_\mathrm{ZPE}=4\pi\kappa N^3T/3$ \eqref{eq:SZPE} instead and
holding the physical radius $R$ fixed as $N\to\infty$ gives
\begin{align}
  S_\mathrm{ZPE}\Big|_{R\ \mathrm{fixed}}
  \xrightarrow{N\to\infty} \frac{8\pi RT}{3}\,N^2,
  \label{eq:SZPE_Rfixed}
\end{align}
genuine $O(N^2)$ scaling, matching the gravity side.
This does not by itself resolve the discrepancy, since no known gravity dual of this
background fixes which of $R$ or $\kappa$ the comparison should use, and
the convexity mismatch of Sec.~\ref{sec:tests} at low temperature is
unaffected by this choice either way; it does show that the power of
$N$ quoted for this system is convention-dependent, not an unambiguous
prediction of the matrix model.
That an $O(N^3)$ entropy can arise from a
Myers/dielectric mechanism at all, independently of this ambiguity, is
supported by the gravity-side literature on the effect: coincident
near-extremal M5-branes have $S\sim N^3T^5$ rather than the $N^2T^3$ of
D3-branes~\cite{klebanov_tseytlin-1996}, and Hatefi, Nurmagambetov and
Park~\cite{hatefi_nurmagambetov_park-2012b,
hatefi_nurmagambetov_park-2012} trace an analogous $N^3$ growth to the
dielectric effect on both the supergravity and Yang--Mills sides of a
D0/D2/D4 system, while showing the same mechanism does not transfer to
D($-1$)/D3 branes. That precedent is only qualitative here --- it
involves different brane content (D0/D4 rather than D0/D2) and a
near-extremal rather than fully thermal comparison --- but it weighs
against reading the present $O(N^3)$ result as symptomatic of an error
in the calculation. This is a separate point from the
convention-dependence discussed above: the M5-brane precedent supports
the physical plausibility of $O(N^3)$ scaling on its own terms,
independently of whether $R$ or $\kappa$ is held fixed in the present
comparison.

The third possibility is that the weak-coupling one-loop approximation 
does not capture the nonperturbative reorganization of degrees of freedom 
expected in the gravity regime.
In particular, higher-loop corrections may modify not only numerical coefficients 
but also the thermodynamic stability of the background through nontrivial resummations 
or phase transitions inaccessible within the Gaussian approximation.
Nothing in the present calculation excludes this, and nothing in it
confirms it either. At present, the available evidence is insufficient
to determine which of these three possibilities is dominant.

Adjudicating between these three causes is the primary task left open
by this work, and three directions follow directly from what has been
computed here. The one-loop truncation can be tested directly by Monte
Carlo simulation of the deformed IKKT matrix model, building on the
complex-Langevin technique already applied to the (undeformed) IKKT
model itself~\cite{anagnostopoulos_azuma_ito_nishimura_okubo_papadoudis-2020},
and in the spirit of the analogous BFSS-model
computation~\cite{hanada-2014}, which would provide a fully non-perturbative
check independent of the loop expansion; a two-loop computation of the
same effective action would isolate the leading correction
analytically. Separately, constructing the finite-temperature,
non-extremal gravity dual of the pure-cubic Myers background --- most
plausibly along the lines of the numerical construction of
Ref.~\cite{costa_greenspan_penedones_santos-2015}, but without the
stabilizing mass deformation --- would settle at once whether the
background choice is responsible, by supplying the missing comparison
target directly. A further, conceptually prior question is how an
intrinsic notion of temperature or horizon could be defined from the
matrix degrees of freedom themselves, independent of any assumed dual.
The present analysis is entirely thermodynamic: the Euclidean
finite-temperature IKKT matrix model has no built-in notion of real,
Lorentzian time. Questions that presuppose one --- infall, quasinormal
ringing, an intrinsic Hawking temperature not borrowed from a dual ---
cannot be posed without first addressing how time is to emerge from the
matrix degrees of freedom, a separate and largely open problem for
IKKT-type constructions in general, pursued to date primarily in the
Lorentzian version of the
model~\cite{kim_nishimura_tsuchiya-2012}. The mass-deformed Euclidean
model faces the issue in its starkest form: its partition function has
neither time nor space built in at all, a point made explicitly in a
recent statistical-mechanical treatment of the polarised IKKT
matrix model~\cite{hartnoll_liu-2025b}.

Taken together, these directions outline a concrete program for
determining whether, and in what sense, black-hole thermodynamics is
realized by the finite-temperature IKKT matrix model. Rather than ruling
out a black-hole interpretation, our results identify the necessary
conditions that any such interpretation must satisfy: the explicit
one-loop expressions for the free energy, entropy, internal energy, and
heat capacity derived here provide a quantitative benchmark against
which future analytical, numerical, and holographic studies can be
tested, 
and any successful proposal for a black-hole sector of the IKKT matrix model 
must reproduce or explain the thermodynamic behavior found here, 
including the entropy scaling under a specified large-$N$ prescription, 
the sign of the heat capacity, and the temperature dependence obtained in this work.
In this sense, the discrepancies found here should be viewed not as
failures of the approach but as concrete theoretical constraints that
future formulations must satisfy.

\section*{Acknowledgments}

The author would like to thank 
the High Energy Accelerator Research Organization (KEK) 
for hosting the workshop ``KEK Theory Workshop 2025'' and
the Yukawa Institute for Theoretical Physics at Kyoto University 
for hosting the workshop ``Strings and Fields 2025'' (YITP-W-25-07). 
Discussions during these meetings were invaluable in developing the ideas presented in this work. 
The author is also grateful to the organizers and speakers of 
the ``KEK Theory Seminar'' and 
the ``YITP Elementary Particle Group Seminar.''

\appendix

\section{Background-field expansion and the one-loop effective action}
\label{app:bgexpansion}

Following the same background-field method as
Ref.~\cite{ishibashi_kawai_kitazawa_tsuchiya-1997}, we record here the
expansion of the deformed IKKT action about a completely general
classical background, and the resulting one-loop effective action, before
specializing to the fuzzy $S^2$ background used in Sec.~\ref{sec:oneloop}.

\subsection{Background-field expansion}

We expand $I=I_B+I_F+I_\mathrm{Myers}$ about a general classical
background $p_\mu$, with bosonic fluctuation $a_\mu$, fermionic
fluctuation $\varphi$ (background fermion $\chi$), following the standard
background-field method,
\begin{align}
  A_\mu = p_\mu+a_\mu, \qquad \psi=\chi+\varphi,
  \label{eq:appsplit}
\end{align}
so that the action organizes as
\begin{align}
  I[p+a,\chi+\varphi] = I^{(0)}[p,\chi]+I^{(1)}[p,\chi;a,\varphi]
  +I^{(2)}[p,\chi;a,\varphi]+\cdots,
  \label{eq:appSexpansion}
\end{align}
order by order in powers of the fluctuations $(a_\mu,\varphi)$. Collecting
terms from $I_B+I_F$,
\begin{align}
  I_B+I_F ={}&
  -\mathrm{Tr}\Big(\frac{1}{4g^2}[p_\mu,p_\nu]^2+\frac{1}{2g^2}\bar\chi\Gamma^\mu[p_\mu,\chi]\Big)
  \notag\\
  &+\frac{1}{g^2}\mathrm{Tr}\,a_\nu\Big([p_\mu,[p^\mu,p^\nu]]+\tfrac12\{\bar\chi\Gamma^\nu,\chi\}\Big)
  -\frac{1}{g^2}\mathrm{Tr}\,\bar\chi\Gamma^\mu[p_\mu,\varphi]
  \notag\\
  &-\frac{1}{g^2}\mathrm{Tr}\Big(\tfrac12[p_\mu,a_\nu]^2-\tfrac12[p_\mu,a^\mu]^2
  +[p_\mu,p_\nu][a^\mu,a^\nu]\notag\\
  &\qquad+\tfrac12\bar\varphi\Gamma^\mu[p_\mu,\varphi]
  +\bar\chi\Gamma^\mu[a_\mu,\varphi]\Big)
  \notag\\
  &-\frac{1}{g^2}\mathrm{Tr}\Big([p_\mu,a_\nu][a^\mu,a^\nu]+\tfrac12\bar\varphi\Gamma^\mu[a_\mu,\varphi]\Big)
  -\frac{1}{4g^2}\mathrm{Tr}[a_\mu,a_\nu]^2,
  \label{eq:appSexpand}
\end{align}
and, similarly, from the Myers term,
\begin{align}
  I_\mathrm{Myers} = \frac{\ii\kappa}{g^2}\varepsilon_{ijk}\,\mathrm{Tr}\Big(
  \tfrac13[p^i,p^j]p^k
  +[p^i,p^j]a^k
  -a^i[p^k,a^j]
  +\tfrac13[a^i,a^j]a^k\Big).
  \label{eq:appSMyersexpand}
\end{align}
Reading off the successive lines of \eqref{eq:appSexpand} together with
the corresponding terms of \eqref{eq:appSMyersexpand}: the first line of
\eqref{eq:appSexpand} and the first term of \eqref{eq:appSMyersexpand}
combine into $I^{(0)}$, the classical action evaluated on the background
alone. The second line of \eqref{eq:appSexpand} and the second term of
\eqref{eq:appSMyersexpand} combine into $I^{(1)}$, linear in the
fluctuations and required to vanish by the classical equations of motion
--- the tadpole condition used in Sec.~\ref{sec:oneloop} to fix the
on-shell background. The third line of \eqref{eq:appSexpand}, together
with the third term of \eqref{eq:appSMyersexpand}, $-\ii\kappa
\varepsilon_{ijk}a^i[p^k,a^j]$, combine into $I^{(2)}$, the quadratic
fluctuation action whose Gaussian integration produces the one-loop
effective action below. The remaining terms are cubic and quartic in the
fluctuations, of higher order in the loop-counting parameter and not
needed at one loop.

We stress in particular that the Myers term contributes not only to
$I^{(0)}$ and $I^{(1)}$ but also this genuine $I^{(2)}$ piece,
$-\ii\kappa\varepsilon_{ijk}\mathrm{Tr}(a^i[p^k,a^j])$, distinct from and
in addition to the $I_B$ cross term $[p_\mu,p_\nu][a^\mu,a^\nu]$ already
present in the third line of \eqref{eq:appSexpand}: on the fuzzy $S^2$
background of Sec.~\ref{sec:oneloop} these two terms cancel each other
exactly, but only once the background satisfies its on-shell condition
$\alpha=\kappa$.

\subsection{Gauge fixing and the quadratic action}

Fixing the residual gauge invariance $\delta A_\mu=\ii[A_\mu,\lambda]$,
$\delta\psi=\ii[\psi,\lambda]$ requires adding the standard
background-covariant gauge-fixing term together with the complex
Grassmann Faddeev--Popov ghosts $b,c$,
\begin{align}
  I_\mathrm{g.f.} = -\frac{1}{2g^2}\mathrm{Tr}[p_\mu,a^\mu]^2, \qquad
  I_\mathrm{F.P.} = \frac{1}{g^2}\mathrm{Tr}\,b\big[p_\mu,[p^\mu+a^\mu,c]\big].
  \label{eq:appSgfFP}
\end{align}
Collecting all terms up to quadratic order in the fluctuations, setting
$\chi=0$ (so the linear term vanishes by the tadpole condition), and
dropping the ghost's own linear-in-$a$ piece (which does not contribute at
one loop), the complete quadratic action entering the path integral is
\begin{align}
  \tilde I^{(2)} = -\frac{1}{g^2}\mathrm{Tr}\Big(\tfrac12[p_\mu,a_\nu]^2
  +[p_\mu,p_\nu][a^\mu,a^\nu]-\ii\kappa\varepsilon_{ijk}a^i[p^k,a^j]\notag\\
  +\tfrac12\bar\varphi\Gamma^\mu[p_\mu,\varphi]+[p_\mu,b][p^\mu,c]\Big).
  \label{eq:appS2tilde}
\end{align}
It is convenient to introduce the adjoint operators $P_\mu$ and $F_{\mu\nu}$
acting on the space of $N\times N$ matrices,
\begin{align}
  P_\mu X \equiv [p_\mu,X], \qquad
  F_{\mu\nu}X \equiv [f_{\mu\nu},X], \qquad f_{\mu\nu}\equiv \ii[p_\mu,p_\nu],
  \label{eq:appPmuFmunu}
\end{align}
in terms of which, after an integration by parts under the trace
($\mathrm{Tr}([p_\mu,X][p_\mu,Y])=-\mathrm{Tr}(X[p_\mu,[p_\mu,Y]])$),
Eq.~\eqref{eq:appS2tilde} becomes
\begin{align}
  \tilde I^{(2)} = \frac{1}{g^2}\mathrm{Tr}\Big(\tfrac12a_\mu(P_\lambda^2\delta_{\mu\nu}-2\ii F_{\mu\nu})a^\nu
  -\ii\kappa\varepsilon_{ijk}a^i[p^k,a^j]
  -\tfrac12\bar\varphi\Gamma^\mu P_\mu\varphi + bP_\lambda^2c\Big),
  \label{eq:appS2adjoint}
\end{align}
in precise agreement with the general background-field result of
Ref.~\cite{ishibashi_kawai_kitazawa_tsuchiya-1997}, up to the additional
Myers term (absent in the pure IKKT matrix model of that reference). This is the
general quadratic action about any classical background $p_\mu$
solving the deformed equations of motion.

\subsection{The one-loop effective action}

Performing the Gaussian integral over $\tilde I^{(2)}$ of
\eqref{eq:appS2adjoint} for a background on which the Myers term happens
to give no residual quadratic contribution (as is the case, after
cancellation against the $I_B$ cross term, for the on-shell fuzzy $S^2$
background of Sec.~\ref{sec:oneloop}), the one-loop effective action
takes the same general form as in
Ref.~\cite{ishibashi_kawai_kitazawa_tsuchiya-1997},
\begin{align}
  \Geff &= -\log\int da\,d\varphi\,dc\,db\;e^{-\tilde I^{(2)}}
  \notag\\
  &{}= \tfrac12\mathrm{Tr}\log(P_\lambda^2\delta_{\mu\nu}-2\ii F_{\mu\nu})
  -\mathrm{Tr}\log(P_\lambda^2)
  -\tfrac14\mathrm{Tr}\log\Big(\big(P_\lambda^2+\tfrac{\ii}{2}F_{\mu\nu}\Gamma^{\mu\nu}\big)
  \big(\tfrac{1+\Gamma_{11}}{2}\big)\Big),
  \label{eq:appGeneralW}
\end{align}
where the three terms are, respectively, the bosonic, ghost, and
fermionic contributions, with $\Gamma^{\mu\nu}\equiv\tfrac12[\Gamma^\mu,
\Gamma^\nu]$ and $\Gamma_{11}\equiv\Gamma^0\Gamma^1\cdots\Gamma^9$ the
ten-dimensional chirality operator projecting onto the physical
Majorana--Weyl fermion. This general formula, applicable to any classical
background satisfying the equations of motion of the deformed IKKT matrix model,
is specialized to the static fuzzy $S^2$ background in
Sec.~\ref{sec:oneloop}, where it is diagonalized explicitly in the basis
of fuzzy spherical harmonics and evaluated as a function of temperature.

\subsection{Diagonalization of the fermionic operator}
\label{app:fermiondiag}

We record here the diagonalization of the fermionic quadratic operator
$(D_F)_r$ on the fuzzy $S^2$ background, used in Sec.~\ref{sec:oneloop}.
Since $(D_F)_r$ is self-adjoint,
$\mathrm{Tr}_{N,S}\ln(D_F)_r=\tfrac12\mathrm{Tr}_{N,S}\ln(D_F)_r^2$, where
$\mathrm{Tr}_{N,S}$ runs over both the $N\times N$ matrix space and the
16-dimensional Majorana--Weyl spinor space. Squaring $(D_F)_r$
and using $(\Gamma^0)^2=\mathbf 1_{16}$, $\{\Gamma^0,\Gamma^i\}=0$, and the
decomposition $\Gamma^i\Gamma^j=\delta^{ij}\mathbf 1+\tfrac12[\Gamma^i,\Gamma^j]$
together with the Jacobi identity for $[L_i,[L_j,\cdot\,]]$,
\begin{align}
  (D_F)_r^2 = (\beta r)^2\mathbf 1+\alpha^2 Q, \qquad
  Q\equiv L^2_\mathrm{ad}+\ii\Sigma_k[L^k,\,\cdot\,], \qquad
  \Sigma_k\equiv \frac14\varepsilon_{kij}[\Gamma^i,\Gamma^j].
\end{align}
Under $\mathrm{SO}(9)\supset \mathrm{SO}(3)\times\mathrm{SO}(6)$, writing
$\Gamma^i=\sigma^i\otimes\mathbf 1_8$ ($i=1,2,3$) for the Pauli matrices
$\sigma^i$ acting on the $\mathrm{SO}(3)$ spinor
($\mathbf{16}_{\mathrm{SO}(9)}=\mathbf 2_{\mathrm{SO}(3)}\otimes\mathbf 8_{\mathrm{SO}(6)}$),
one finds $\Sigma_k=2\ii\tilde S_k$ with $\tilde S_k\equiv\sigma^k/2\otimes\mathbf1_8$
the normalized $\mathrm{SO}(3)$ spin generator, $\tilde S^2=\tfrac34\mathbf1_{16}$, so
that $Q=L^2_\mathrm{ad}-2\tilde S_k L^{k(\mathrm{ad})}$. Since $L_k^{(\mathrm{ad})}$
and $\tilde S_k$ act on different tensor factors (the matrix and spinor
indices, respectively) and each separately satisfies the
$\mathrm{SU}(2)$ algebra, $J_k\equiv L_k^{(\mathrm{ad})}+\tilde S_k$
closes into a genuine total angular momentum,
$[J_i,J_j]=\ii\varepsilon_{ijk}J^k$. Expanding
$J^2=L^2_\mathrm{ad}+\tilde S^2+2\tilde S_kL^{k(\mathrm{ad})}$ and solving
for the spin--orbit term,
\begin{align}
  Q = L^2_\mathrm{ad}-\Big(J^2-L^2_\mathrm{ad}-\tfrac34\Big)
  = 2L^2_\mathrm{ad}-J^2+\tfrac34\mathbf1_{16}, \qquad
  Q\big|_{j} = 2l(l+1)-j(j+1)+\tfrac34.
\end{align}
Combining the orbital angular momentum $l$ of $\hat Y_{l\mu}$ with the spin
$\tilde s=\tfrac12$ of $\tilde S_k$ via $j=l\pm\tfrac12$
(Clebsch--Gordan decomposition) gives two towers of eigenvalues,
\begin{align}
  \mu^{(+)}_{r,l} &= (\beta r)^2+\alpha^2 l^2, \qquad \text{degeneracy } 16(l+1)
  \quad (l\geq0), \label{eq:appmup}\\
  \mu^{(-)}_{r,l} &= (\beta r)^2+\alpha^2(l+1)^2, \qquad \text{degeneracy } 16l
  \quad (l\geq1). \label{eq:appmum}
\end{align}
(the $j=l+\tfrac12$ branch, of dimension $2(l+1)$ per copy of the
8-fold-degenerate spin-$\tfrac12$ space, gives $\mu^{(+)}_{r,l}$; the
$j=l-\tfrac12$ branch, of dimension $2l$, gives $\mu^{(-)}_{r,l}$). The
tower $\mu^{(+)}_{r,l}$ is defined for $l\geq0$ and $\mu^{(-)}_{r,l}$ for
$l\geq1$; the only member with vanishing frequency at $r=0$ is
$\mu^{(+)}_{r,0}=(\beta r)^2$ (the $l=0$, $j=\tfrac12$ mode), which is a
$\mathrm{U}(1)^N$ gauge-redundant zero mode and is excluded, exactly as for the
$l=0$ boson. The correctly-diagonalized spectrum above has no further zero at $l=1$
in the ``$(-)$'' tower: $\mu^{(-)}_{r,1}\big|_{r=0}=4\alpha^2\neq0$, so
exactly one gauge-redundant zero mode occurs per sector, matching the
single bosonic trace mode one-to-one.

\section{Details of the Matsubara sum}
\label{app:matsubara}

This appendix gives the detailed derivation of the Matsubara-frequency
sums
\begin{align}
  \sum_{n\in\mathbb{Z}}\ln\big[(\beta n)^2+\omega^2\big], \qquad
  \sum_{r\in\mathbb{Z}+\frac12}\ln\big[(\beta r)^2+\omega^2\big]
  \label{eq:appMatsubaraTargets}
\end{align}
quoted in Sec.~\ref{sec:matsubara}, by two independent methods: a residue
(contour-integral) evaluation, and a $\zeta$-function regularization that
additionally fixes the additive constant left undetermined by the residue
method.

\subsection{Residue evaluation}

For the bosonic sum, differentiating with respect to $\omega^2$,
\begin{align}
  \frac{\partial}{\partial(\omega^2)}\sum_{n\in\mathbb{Z}}
  \ln\big[(\beta n)^2+\omega^2\big]
  = \sum_{n\in\mathbb{Z}}\frac{1}{(\beta n)^2+\omega^2}.
  \label{eq:appBderiv}
\end{align}
Using the standard residue formula $\sum_{n\in\mathbb{Z}}f(n)=
-\sum_{\text{poles of }f}\mathrm{Res}[\pi\cot(\pi z)f(z)]$ with
$f(z)=1/(\beta^2z^2+\omega^2)$, whose poles are at $z=\pm\ii\omega/\beta$,
and $\cot(\ii\theta)=-\ii\coth\theta$,
\begin{align}
  \mathrm{Res}_{z=\ii\omega/\beta}\big[\pi\cot(\pi z)f(z)\big]
  = \frac{-\pi\coth(\pi\omega/\beta)}{2\beta\omega},
\end{align}
with an equal contribution from the pole at $z=-\ii\omega/\beta$ by
symmetry. Hence
\begin{align}
  \sum_{n\in\mathbb{Z}}\frac{1}{(\beta n)^2+\omega^2}
  = \frac{\pi\coth(\pi\omega/\beta)}{\beta\omega}.
  \label{eq:appBresidue}
\end{align}
Integrating \eqref{eq:appBresidue} over $\omega^2$ (equivalently, using
$\int\coth(ax)\,\mathrm{d}x=\tfrac1a\ln\sinh(ax)$ and
$\sinh(x)=\tfrac{e^x}2(1-e^{-2x})$), and dropping the divergent lower
endpoint into an undetermined additive constant $\tilde C(\beta)$,
\begin{align}
  \sum_{n\in\mathbb{Z}}\ln\big[(\beta n)^2+\omega^2\big]
  = \frac{2\pi\omega}{\beta}+2\ln\big(1-e^{-2\pi\omega/\beta}\big)+\tilde C(\beta).
  \label{eq:appBsumresidue}
\end{align}

For the fermionic sum, the half-integer analogue of the residue formula
is $\sum_n f(n+\tfrac12)=+\sum_{\text{poles}}\mathrm{Res}[\pi\tan(\pi z)
f(z)]$ (the relative plus sign, compared to the bosonic formula's minus,
follows from $\mathrm{Res}[\pi\tan(\pi z),n+\tfrac12]=-1$ rather than
the bosonic $\mathrm{Res}[\pi\cot(\pi z),n]=+1$); using
$\tan(\ii\theta)=\ii\tanh\theta$, both poles $z=\pm\ii
\omega/\beta$ contribute equally, giving
\begin{align}
  \sum_{r\in\mathbb{Z}+\frac12}\frac{1}{(\beta r)^2+\omega^2}
  = \frac{\pi\tanh(\pi\omega/\beta)}{\beta\omega},
  \label{eq:appFresidue}
\end{align}
and, integrating as above with $\int\tanh(ax)\,\mathrm{d}x=\tfrac1a
\ln\cosh(ax)$ and $\cosh(x)=\tfrac{e^x}2(1+e^{-2x})$,
\begin{align}
  \sum_{r\in\mathbb{Z}+\frac12}\ln\big[(\beta r)^2+\omega^2\big]
  = \frac{2\pi\omega}{\beta}+2\ln\big(1+e^{-2\pi\omega/\beta}\big)+\tilde C_F(\beta).
  \label{eq:appFsumresidue}
\end{align}
In both \eqref{eq:appBsumresidue} and \eqref{eq:appFsumresidue}, the first
term is the zero-point energy and the second is the thermal
(Bose--Einstein or Fermi--Dirac) occupation contribution; the residue
method is insensitive to any $\omega$-independent additive constant,
leaving $\tilde C(\beta),\tilde C_F(\beta)$ undetermined.

\subsection{\texorpdfstring{$\zeta$}{zeta}-function cross-check}
\label{app:zetacrosscheck}

The additive constants can be fixed by an independent derivation via
spectral zeta functions,
\begin{align}
  \zeta_B(s;\omega,\beta) \equiv \sum_{n\in\mathbb{Z}}\big[(\beta n)^2+\omega^2\big]^{-s},
  \qquad
  \sum_n\ln[\cdots] \equiv -\zeta_B'(0),
  \label{eq:appzetadef}
\end{align}
and similarly for $\zeta_F(s;\omega,\beta)$ with $r\in\mathbb{Z}+\tfrac12$.
Using the Mellin representation $\Gamma(s)A^{-s}=\int_0^\infty\mathrm dt\,
t^{s-1}e^{-tA}$ and the Poisson resummation of the Jacobi theta function,
\begin{align}
  \sum_{n\in\mathbb{Z}}e^{-t\beta^2n^2} = \sqrt{\frac{\pi}{t\beta^2}}
  \sum_{k\in\mathbb{Z}}e^{-\pi^2k^2/(\beta^2t)},
  \label{eq:apppoissonboson}
\end{align}
(with an extra alternating sign $(-1)^k$ on the right-hand side for the
half-integer, fermionic lattice), one separates the $k=0$ term from
$k\neq0$, writing $\zeta_B(s;\omega,\beta)=f(s)+g(s)$ with
\begin{align}
  f(s) \equiv{}& \frac{\sqrt\pi\,\Gamma(s-\tfrac12)}{\beta\,\Gamma(s)}\,\omega^{1-2s}, \qquad
  g(s) \equiv \frac{4\sqrt\pi}{\beta\Gamma(s)}\sum_{k=1}^\infty
  \Big(\frac{\pi k}{\beta\omega}\Big)^{s-\frac12}K_{s-\frac12}\!\Big(\frac{2\pi k\omega}{\beta}\Big),
  \label{eq:appfgdef}
\end{align}
where $K_\nu$ is the modified Bessel function, obtained via the standard
integral
\begin{align}
  \int_0^\infty\mathrm dt\,t^{\nu-1}e^{-at-b/t}=2(b/a)^{\nu/2}K_\nu(2\sqrt{ab}).
\end{align}
Near $s=0$, using $1/\Gamma(s)=s+O(s^2)$ and
$\Gamma(s-\tfrac12)=-2\sqrt\pi+O(s)$, together with the closed form
$K_{-1/2}(x)=\sqrt{\pi/(2x)}\,e^{-x}$ and
$\sum_{k=1}^\infty x^k/k=-\ln(1-x)$,
\begin{align}
  f(0)=0,&\qquad f'(0)=-\frac{2\pi\omega}{\beta},
  \notag\\
  g(0)=0,&\qquad g'(0)=-2\ln\big(1-e^{-2\pi\omega/\beta}\big).
  \label{eq:appfprime0}
\end{align}
Hence $\zeta_B(0)=f(0)+g(0)=0$ --- so the regularized sum carries no
dependence on an arbitrary renormalization scale --- and
\begin{align}
  \sum_{n\in\mathbb{Z}}\ln\big[(\beta n)^2+\omega^2\big]
  = \frac{2\pi\omega}{\beta}+2\ln\big(1-e^{-2\pi\omega/\beta}\big),
  \label{eq:appbosonzetafinal}
\end{align}
reproducing \eqref{eq:appBsumresidue} exactly and fixing
$\tilde C(\beta)=0$. The fermionic case proceeds identically, with the
$k\neq0$ term of \eqref{eq:appfgdef} replaced by its alternating-sign
counterpart ($\sum_k(-1)^kx^k/k=-\ln(1+x)$ in place of
$\sum_kx^k/k=-\ln(1-x)$), giving $\zeta_F(0)=0$ and
\begin{align}
  \sum_{r\in\mathbb{Z}+\frac12}\ln\big[(\beta r)^2+\omega^2\big]
  = \frac{2\pi\omega}{\beta}+2\ln\big(1+e^{-2\pi\omega/\beta}\big),
  \label{eq:appfermionzetafinal}
\end{align}
reproducing \eqref{eq:appFsumresidue} exactly and fixing
$\tilde C_F(\beta)=0$. The two independent methods --- residue and
$\zeta$-function --- thus agree completely on the $\omega$-dependent
content, with the $\zeta$-function regularization additionally fixing the
previously free additive constants to zero, as used in
Sec.~\ref{sec:matsubara}.

\subsection{Application to the bosonic, ghost, and fermionic towers}

Applying \eqref{eq:appbosonzetafinal} with $\omega=\Omega_l=\alpha\sqrt{l(l+1)}$
to the combined boson-plus-ghost sum \eqref{eq:BGh2_paper},
\begin{align}
  W^{(1)}_B+W^{(1)}_\mathrm{gh}
  = \tfrac72\sum_{l=1}^{N-1}(2l+1)\left[\frac{2\pi\Omega_l}{\beta}
  +2\ln\big(1-e^{-2\pi\Omega_l/\beta}\big)\right].
  \label{eq:appBghfinal}
\end{align}
Applying \eqref{eq:appfermionzetafinal} with $\omega=\Omega_l^{(+)}=
\alpha l$ and $\omega=\Omega_l^{(-)}=\alpha(l+1)$
to the fermionic sum \eqref{eq:Gamma1F_paper},
\begin{align}
  W^{(1)}_F = -4\sum_{l=1}^{N-1}\Bigg[&(l+1)\left(\frac{2\pi\Omega_l^{(+)}}{\beta}
  +2\ln\Big(1+e^{-2\pi\Omega_l^{(+)}/\beta}\Big)\right)
  \notag\\
  &+l\left(\frac{2\pi\Omega_l^{(-)}}{\beta}+2\ln\Big(1+e^{-2\pi\Omega_l^{(-)}/\beta}\Big)\right)\Bigg].
  \label{eq:appFfinal}
\end{align}
Both $l$-sums in \eqref{eq:appBghfinal} and \eqref{eq:appFfinal} run over
the full range $l=1,\ldots,N-1$, with every frequency
$\Omega_l,\Omega_l^{(+)},\Omega_l^{(-)}$ strictly positive throughout
this range (App.~\ref{app:fermiondiag}), so no further mode requires
special treatment beyond the single excluded $l=0$ gauge zero mode
already accounted for in \eqref{eq:BGh2_paper} and
\eqref{eq:Gamma1F_paper}. Adding \eqref{eq:appBghfinal} and
\eqref{eq:appFfinal} reproduces
the split \eqref{eq:Gamma1split_paper}--\eqref{eq:Tl_paper} of
Sec.~\ref{sec:oneloop} into zero-point and thermal pieces.


\end{document}